\documentclass[preprint,5p,times,twocolumn]{elsarticle}

\usepackage{amssymb}
\usepackage{appendix}
\usepackage{amsmath}
\usepackage[misc]{ifsym}
\usepackage{multirow}
\usepackage{float}
\usepackage{multicol}
\usepackage{soul}
\usepackage{color, xcolor}
\usepackage[colorlinks=true, linkcolor=blue, citecolor=blue, urlcolor=blue]{hyperref}

\journal{Knowledge-Based Systems}

\begin{document}

\begin{frontmatter}


\title{Unsupervised Social Event Detection via Hybrid Graph Contrastive Learning and Reinforced Incremental Clustering}
\author{Yuanyuan Guo\fnref{label1,label2}}
\address[label1]{Science \& Technology on Integrated Information System Laboratory, Institute of Software, Chinese Academy of Sciences, Beijing, China. \fnref{label1}}
\address[label2]{University of Chinese Academy of Sciences, Beijing, China. \fnref{label2}}
\address[label3]{National Defense University, Beijing, China. \fnref{label3}}
\author{Zehua Zang\fnref{label1,label2}}
\author{Hang Gao\fnref{label1,label2}}
\author{Xiao Xu\fnref{label3}}
\author{Rui Wang\fnref{label1,label2}}
\author{Lixiang Liu\fnref{label1,label2}}
\author{\Letter Jiangmeng Li\fnref{label1,label2}}
\ead{jiangmeng2019@iscas.ac.cn}

\begin{abstract}
Detecting events from social media data streams is gradually attracting researchers. The innate challenge for detecting events is to extract discriminative information from social media data thereby assigning the data into different events. Due to the excessive diversity and high updating frequency of social data, using supervised approaches to detect events from social messages is hardly achieved. To this end, recent works explore learning discriminative information from social messages by leveraging graph contrastive learning (GCL) and embedding clustering in an unsupervised manner. However, two intrinsic issues exist in benchmark methods: conventional GCL can only roughly explore partial attributes, thereby insufficiently learning the discriminative information of social messages; for benchmark methods, the learned embeddings are clustered in the latent space by taking advantage of certain specific prior knowledge, which conflicts with the principle of unsupervised learning paradigm. In this paper, we propose a novel unsupervised social media event detection method via hybrid graph contrastive learning and reinforced incremental clustering (HCRC), which uses hybrid graph contrastive learning to comprehensively learn semantic and structural discriminative information from social messages and reinforced incremental clustering to perform efficient clustering in a solidly unsupervised manner. We conduct comprehensive experiments to evaluate HCRC on the Twitter and Maven datasets. The experimental results demonstrate that our approach yields consistent significant performance boosts. In traditional incremental setting, semi-supervised incremental setting and solidly unsupervised setting, the model performance has achieved maximum improvements of 53\%, 45\%, and 37\%, respectively. 
\end{abstract}
\onecolumn







\begin{keyword}
Event detection \sep Unsupervised learning \sep Graph contrastive learning \sep Incremental clustering \sep Reinforcement learning
\end{keyword}

\end{frontmatter} 


\begin{figure*}[t]
\begin{center}
\includegraphics[width=1.0\textwidth]{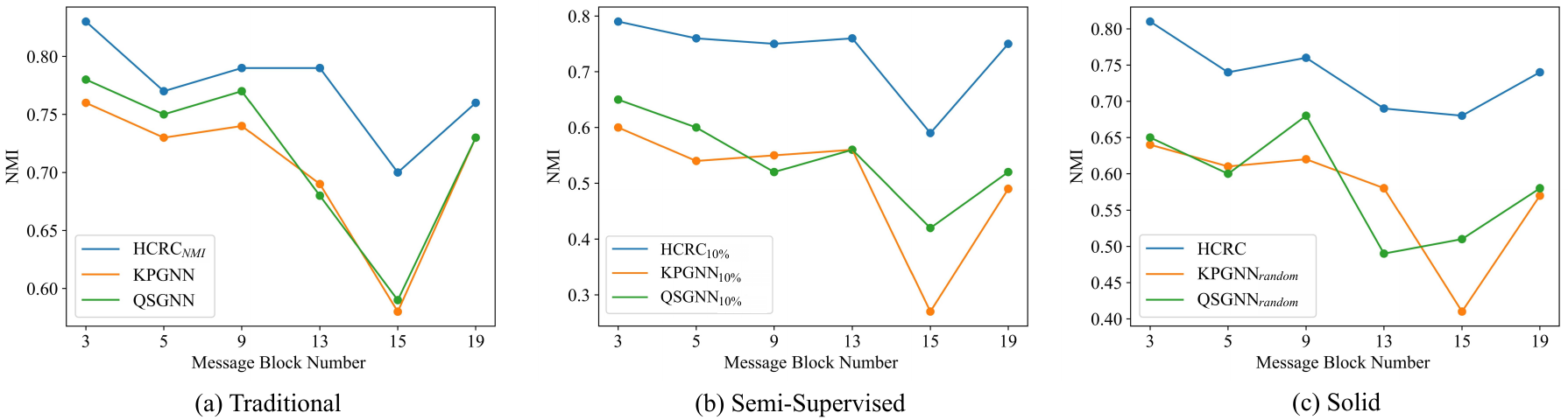}
\caption{ Comparison of various settings using NMI Metric. Experimental results show our method outperforms the baselines under different settings (refer to Section \ref{sec:increexp} for details).}
\label{fig1}
\end{center}

\end{figure*}
\section{Introduction} \label{intro}
With the continuous development of social networking services, the rapidly growing users spread worldwide. According to statistics, there are 4.74 billion social media users around the world, equating to 59.3\% of the total global population \cite{datareportal2022}. Social media becomes a focal point for researchers to gather information on events happening immediately \cite{ritter2012open}. A long-lasting challenge of such behavior is countless routine instant messages on social media. For instance, around 10,033 tweets are posted per second on average as of May 2022 \cite{yaqub2022}, and the events of such messages are generally inconsistent with historical data so that classifying the instant events based solely on historical data is inaccessible and labeling enough new events requires inconceivable efforts. However, the previous works \cite{mikolov2013efficient,blei2003latent,kusner2015word,devlin2018bert,graves2005framewise} primarily focused on static event detection, which clearly does not align with real-world application scenarios. Therefore, recent works explore capturing discriminative information from instant events thereby performing the incremental clustering by leveraging graph neural networks (GNNs) and contrastive learning in an unsupervised manner \cite{KPGNN}. 

In the realm of unsupervised social event detection, one engages in the intricate process of unearthing clusters that embody real-world events within the ever-flowing social stream (refer to Section \ref{section_problem} for details). State-of-the-art unsupervised social event detection methods \cite{peng2019fine,liu2020story,KPGNN,QSGNN} explore to jointly learn semantic and structural information from the social data by leveraging GNNs. Specifically, the content and corresponding attributes, e.g., location, post time, etc., of social messages are mapped from the data space into the latent space by a fixed pre-trained feature extractor, e.g., en\_core\_web\_lg \cite{spacy}. 
Benchmark methods explore the relationships between social messages and further extract discriminative information from the raw data to convert it into graph-based data, i.e., nodes of the graph denote the social messages, thereby assigning the data into different events. The semantic and structural information can be jointly captured by leveraging a well-designed GNN and further contrasting the node embeddings. However, such a learning paradigm cannot sufficiently explore the semantic information of social messages since only partial attributes are considered by the model, and the semantic information is learned by a fixed feature extractor. In the latent space, the learned embeddings of social messages are clustered. Yet, there exists an intrinsic issue with current approaches. In detail, the adopted clustering approach requires specific prior knowledge, e.g., the pre-set hyperparameter $k$ for K-Means, which conflicts with the principle of the unsupervised learning paradigm.

To this end, we propose HCRC, short for \textit{\textbf{H}ybrid graph \textbf{C}ontrastive learning and \textbf{R}einforced incremental \textbf{C}lustering}, which is orthogonal to existing methods in two key ingredients: 1) HCRC innovatively proposes a simple yet effective approach to build social message graphs comprehensively including the content and attributes, and the proposed hybrid graph contrastive learning contains the graph-level and node-level contrasts, which jointly empowers the model to sufficiently learn the semantic and structural information from the social data. The graph-level contrast builds a trainable approach to learning discriminative semantic information from the content and attributes of social messages, and the node-level contrast improves the model to capture valuable structural information from the social message graph; 2) The proposed reinforced incremental clustering enables HCRC to perform efficient clustering on the instantly updated social data in a solidly unsupervised manner, which is proved in Fig.~\ref{fig1}. Concretely, the contributions of this paper are four-fold: 

\begin{itemize}
	\item  We present a novel unsupervised social event detection architecture, namely HCRC, and empirically demonstrate the effectiveness of HCRC on various benchmarks.
	\item We propose a simple yet effective approach to building social message graphs, and the proposed hybrid graph contrastive learning boosts the model's capacity to learn discriminative social message embeddings.
	\item Guided by deep reinforcement learning, a density-based spatial clustering module is proposed to perform incremental social event detection in a solidly unsupervised manner.
	\item Sufficient experiments further prove the interpretability and effectiveness of the proposed HCRC.
\end{itemize} 

\section{Related Works}

\subsection{Social Event Detection}
An event is an occurrence causing a change in the volume of text data that discusses the associated topic at a specific time \cite{event_work1}. Social event detection aims at clustering social messages based on their correlations from social media streams.
Some classic works \cite{event_work2,event_work3} design different feature engineering to build social message objects. Later, more works \cite{event_work4,event_work5,event_work6} adopt pre-trained language models to get better representations of social messages. To better model relationships between messages, KPGNN \cite{KPGNN} first uses a heterogeneous GNN-based knowledge-preserving incremental social event detection model.In order to dynamically adjust to the evolving data, KPGNN incorporates contrastive loss terms that effectively handle varying numbers of event classes \cite{KPGNN}. QSGNN \cite{QSGNN} enhances the transfer of knowledge from known to unknown domains by leveraging the most valuable information from known samples and reliable knowledge transfer techniques. Researchers also detect events and discover event evolution in heterogeneous information graphs \cite{event_work7}. Due to the ever-changing nature of social media, some works focus on dynamic representations of heterogeneous information graphs \cite{event_work8,event_work9,event_work10}. 

\subsection{Graph Contrastive Learning}
A graph contrastive learning (GCL) framework usually consists of a graph views generation component to construct positive and negative views and a contrastive objective to discriminate positive pairs from negative pairs \cite{gcl_work1}. Grace \cite{gcl_work2} generates two graph views by corruption and learns node representation by maximizing the agreement of node representations in these two views. Further, ProGCL \cite{gcl_work7} constitutes a measure for negatives’ hardness together with similarity to tackle the problem of hard negative samples. Several works have proposed trainable augmentation strategies \cite{gcl_work4,gcl_work5} to learn a drop probability distribution over nodes or edges. Differently, SimGRACE \cite{gcl_work6} proposes a Simple framework for GCL, which does not require data augmentations.

\subsection{Incremental Clustering Algorithm}
An incremental algorithm can process its input serially, i.e., in the order that the input is fed to the algorithm, unlike an offline algorithm with the entire input available from the start. For example, using the hash strategy and avoiding much similarity calculation, Locality-Sensitive Hashing (LSH) \cite{anand2011mining} is widely used for data clustering and nearest neighbor search. SinglePass clustering \cite{papka1998line} is a simple and efficient incremental clustering algorithm. Since each data only needs to flow through the algorithm once, the efficiency is much higher than offline algorithms such as K-Means \cite{jain1988algorithms} or KNN \cite{fix1989discriminatory}. There are also works that improve other offline algorithms to incremental scenarios, such as incremental K-Means and incremental DBSCAN algorithms \cite{chakraborty2014performance}. Some recent works focus on dynamically adjusting algorithms to better adapt to streaming data. 

\subsection{Deep Reinforcement Learning}
Deep Reinforcement Learning (DRL) is learning an agent making sequential decisions to maximize accumulative rewards. There are two categories of DRL methods: value-based and policy-gradient methods. The value-based methods \cite{dqn,doubledqn,duelingdqn,rainbow} are limited to the environments with discrete action space estimating the Q-value of the actions and choosing the largest one. By comparison, the policy-gradient methods \cite{dpg,ddpg,a2c,ppo,sac} are designed to work with environments that have either discrete or continuous action spaces, and they use action distributions, such as Normal distribution, to sample actions during the learning process.
\begin{figure*}[t]
	\vskip -0in
	\begin{center}
		\centerline{\includegraphics[width=0.7\textwidth]{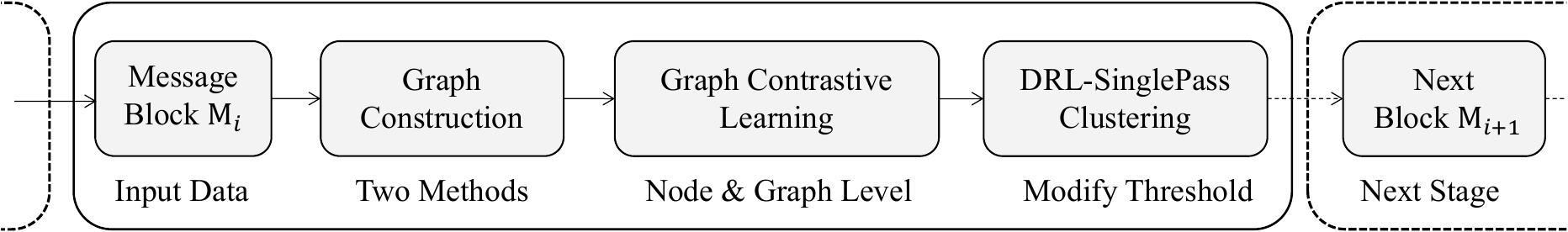}}
		\vskip -0.in
		\caption{Pipeline of the proposed HCRC.}
		\label{fig:pipeline}
	\end{center}
\vspace{-0.5em}
\end{figure*}

\section{Preliminary}
\subsection{Graph Convolutional Network}
GNNs \cite{gori2005new,velivckovic2017graph,kipf2016semi,xu2018powerful} have received much attention recently. Specifically, graph convolutional network (GCN) \cite{kipf2016semi} is widely used due to its excellent ability to analyze graph-structured data. In detail, the architecture of GCN is defined as:
\begin{equation}
\textbf{H}^{(l+1)} =\sigma(\hat{\textbf{D}}^{-1/2}\hat{\textbf{A}}\hat{\textbf{D}}^{-1/2}\textbf{H}^{(l)}\textbf{W}^{(l)}),
\label{eq:gcn}
\end{equation}
where $\textbf{H}^{(l)}$ is the node embedding matrix of the \emph{l}-th layer for \emph{l} $\in$ [1,\ldots, L]. $\hat{\textbf{A}}$ is the adjacency matrix with self-loops. $\hat{\textbf{D}}$ is the degree matrix and $\hat{\textbf{D}_{ii}}=\sum_{j} \hat{\textbf{A}}_{ij}$. $\textbf{W}^{(l)}$ is the trainable weight matrix of the \emph{l}-th layer. $\sigma(\cdot)$ is an activation function, e.g., $\rm{ReLU}(\cdot)=\rm{max}(0, \cdot)$.

\subsection{Graph Contrastive Learning}

Recently, graph contrastive learning (GCL) has emerged as a promising approach to learning graph representations. The primary objective of GCL is to facilitate the creation of highly effective representations through the agglomeration of semantically similar pairs and the divergence of dissimilar pairs. For a given graph $\mathcal{G}$, two graph views, $\hat{\mathcal{G}}_1$ and $\hat{\mathcal{G}}_2$, are generated via augmentations $\mathcal{T}(\cdot)$, which consist of node dropping, edge perturbation, attribute masking, and subgraph \cite{you2020graph}. Then, a GNN-based encoder, denoted as $f(\cdot)$, extracts node representations $\textbf{U}$ and $\textbf{V}$ for different views. Specifically, node embeddings in the two generated views are denoted as $\textbf{U}=f\left(\textbf{X}_1, \textbf{A}_1\right)$ and $\textbf{V}=f\left(\textbf{X}_2, \textbf{A}_2\right)$, where $\textbf{X}_*$ and $\textbf{A}_*$ are the feature matrices and adjacency matrices of the views. After that, a contrastive objective is employed to contrast the embeddings of the same node in the two views with other node embeddings. Specifically, for any node $k$, the embedding obtained in one view $\textbf{u}_k$ is deemed as the anchor, and the embedding of it in the other view $\textbf{v}_k$ is regarded as the positive sample, whereas the remaining embeddings in two views are considered negative samples.
Referring to the loss proposed in GCA  \cite{zhu2021graph}, we define the pairwise objective for each positive pair $\left(\textbf{u}_k, \textbf{v}_k\right)$ as
\begin{equation}
\label{eq:gcaloss}
\small
\begin{aligned}
& \ell\left(\textbf{u}_k, \textbf{v}_k\right)=  \log \frac{e^{\theta\left(\textbf{u}_k, \textbf{v}_k\right) / \tau}}{e^{\theta\left(\textbf{u}_k, \textbf{v}_k\right) / \tau}+\sum\limits_{m\neq k} e^{\theta\left(\textbf{u}_k, \textbf{v}_m\right) / \tau}+\sum\limits_{m\neq k} e^{\theta\left(\textbf{u}_k, \textbf{u}_m\right) / \tau}},
\end{aligned}
\end{equation}
where $\tau$ is a temperature parameter. $\textbf{u}_*\in \textbf{U}$ and $\textbf{v}_*\in \textbf{V}$. $\theta(\textbf{u}, \textbf{v})=$ $s(g(\textbf{u}), g(\textbf{v}))$, where $s(\cdot, \cdot)$ is the cosine similarity and $g(\cdot)$ is the nonlinear projection, which is a two-layer perception model. The objective to be maximized is defined as the average over all positive pairs given by
\begin{equation}
\label{eq:gcaloss2}
    \mathcal{J}=\frac{1}{2 N} \sum_{k=1}^N\left[\ell\left(\textbf{u}_k, \textbf{v}_k\right)+\ell\left(\textbf{v}_k, \textbf{u}_k\right)\right].
\end{equation}

\subsection{Reinforcement Learning}

\textit{Concepts}. The state space denoted as $\mathcal{S}$ represents the agent's current situation, which is treated as the input of agents. The action space, denoted as $\mathcal{A}$, is a set of candidate actions for the agent. The reward function denoted as $\mathcal{R}$, trains agents with respect to maximizing the cumulative reward.

\textit{Trajectory}. For reinforcement learning, at $t$-th step, the agent gets the state $s_t$ from the environment and then samples an action from its policy $\textbf{a}_t\sim \pi(\textbf{a}_t|\textbf{s}_t)$, where $0\leq t \leq T$. The executed action leads the environment to a new state $\textbf{s}_{t+1} \sim p(\textbf{s}_t|\textbf{a}_{t+1})$, and the agent gets a new reward $\textbf{r}_t=r(\textbf{s}_t,\textbf{a}_t,\textbf{s}_{t+1})$ where $p$ is the dynamics of the environment, and $r$ is the reward function. The trajectory $\tau$ is represented as $\{\textbf{s}_0,\textbf{a}_0,\textbf{r}_0,...,\textbf{s}_{T-1},\textbf{a}_{T-1},\textbf{r}_{T-1},\textbf{s}_{T}\}$.

\section{Methodology}

\begin{figure}[t]
	\vskip -0in
	\begin{center}
\centerline{\includegraphics[width=0.35\textwidth]{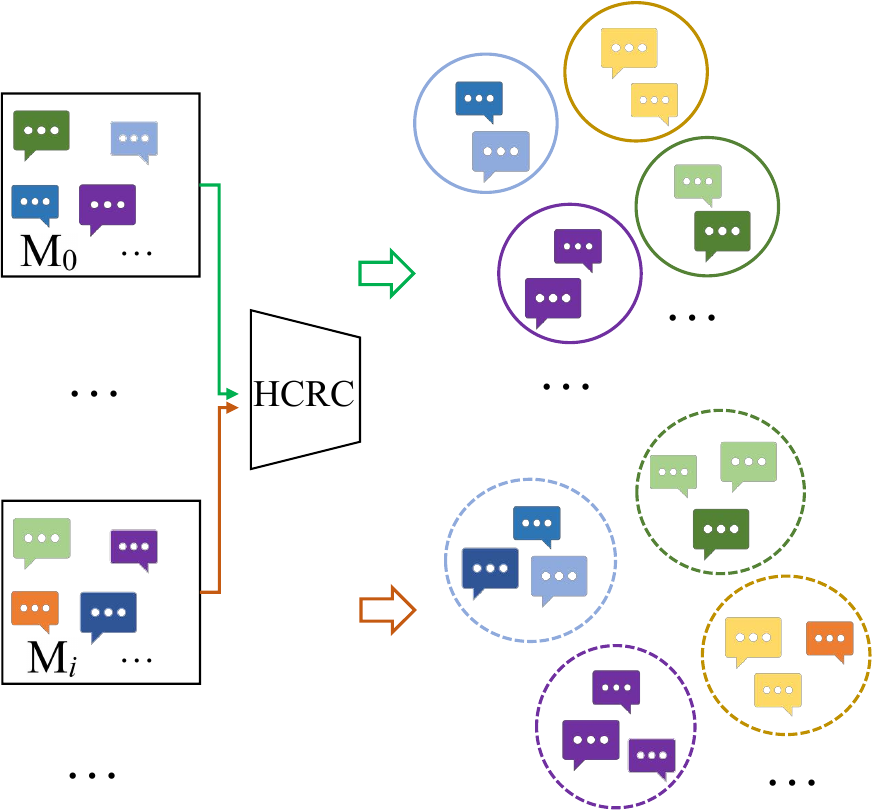}}
		\caption{The visual representation of proposed problem.} 
        \label{fig:problem}
	\end{center}

\end{figure}

\begin{figure*}[t]
	\vskip -0in
	\begin{center}
		\centerline{\includegraphics[width=1\textwidth]{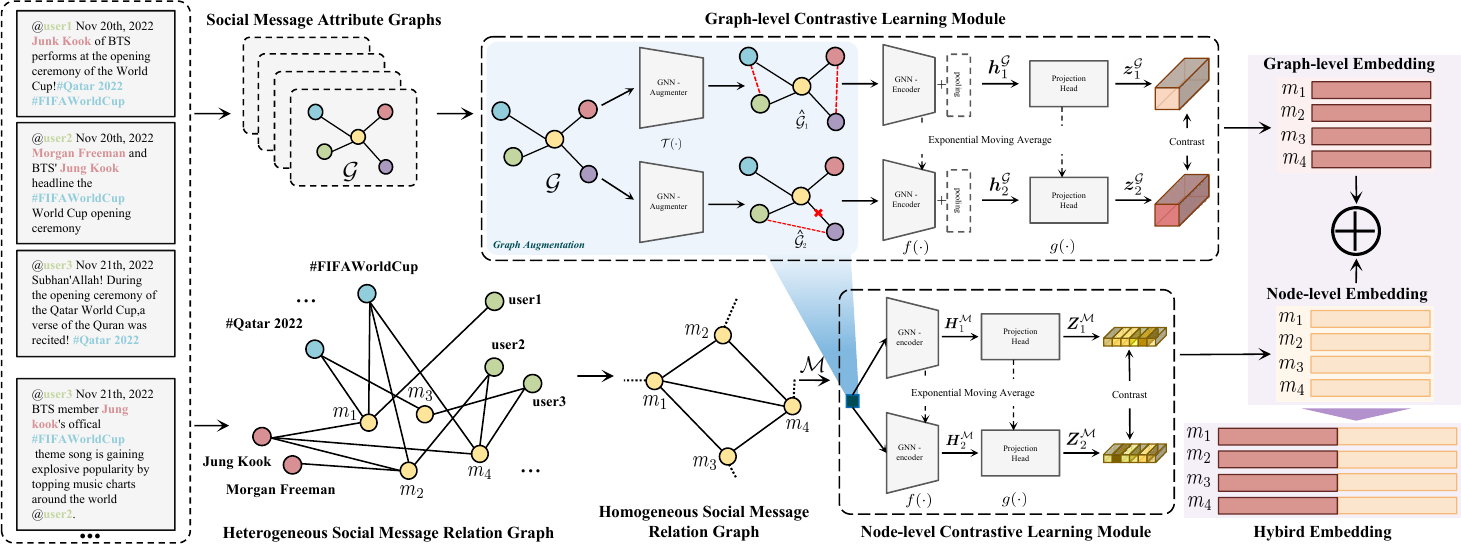}}
		\vskip -0.in
        \caption{Details of the Graph Construction and Hybrid Graph Contrastive Learning.}
		\label{fig:algoframe}
	\end{center}

\end{figure*}
\begin{table}[]
\small
\renewcommand\arraystretch{0.95}
\centering
\caption{Glossary of Notations.}\label{notations}
\begin{tabular}{c|c}
\hline
Notation    & Description                                            \\ \hline
M$_i$      & Message block $i$                                     \\
m$_i$      & A message                                           \\
$\textbf{X}_*$       & The feature matrices                             \\
$\textbf{A}_*$       & The adjacency matrices                            \\
$\mathcal{S}$        & State space                                         \\
$\mathcal{A}$        & Action space                                        \\
$\mathcal{R}$        & Reward function                                     \\
$\mathcal{G}$        & Social message attribute graph                      \\
$\mathcal{M}$        & Social message relation graph                       \\
$\hat{\mathcal{G}}_*$ & Augmented social message attribute graph            \\
$\hat{\mathcal{M}}_*$ & Augmented social message relation graph             \\
$\boldsymbol{h}_*^{\mathcal{G}}$      & Global representation of $\hat{\mathcal{G}}_*$ \\
$\boldsymbol{z}_*^{\mathcal{G}}$        & $\boldsymbol{h}_*^{\mathcal{G}}$  through projection head       \\
$\boldsymbol{H}_*^{\mathcal{M}}$          & The node representations of $\mathcal{M}$     \\
$\boldsymbol{Z}_*^{\mathcal{M}}$         & $\boldsymbol{H}_*^{\mathcal{M}}$     through projection head        \\
f($\cdot$)       & GNN-encoder                                         \\
g($\cdot$)       & Projection head                                     \\
$\epsilon_t$         & The current minimum neighbor distance               \\
$\operatorname{coh}_t$       & The average cohesion distance                       \\
$\operatorname{sep}_t$       & The average separation distance                     \\
$DI_t$        &  Dunn Index                                      \\
$S_t$           & Silhouette coefficient                          \\
$V^W_t$        & The overall within-cluster variance                 \\
$V^B_t$        & The overall between-cluster variance                \\ \hline
\end{tabular}

\end{table}

\subsection{Notations}
We summarize the main notations in Table \ref{notations}.

\subsection{Problem Formulation}
\label{section_problem}
The social stream is a continuous sequence of messages. $\mathrm{M}_i$ denotes a message block containing all the messages during a certain time period and $\mathrm{M}_i=\left\{m_j|1 \leq j \leq| \mathrm{M}_i\mid\right\}$, where $\left|\mathrm{M}_i\right|$ is the total number of messages contained by $\mathrm{M}_i$, and $m_j$ is a specific message. As shown in Fig. \ref{fig:problem}, given a message block $\mathrm{M}_i$, a social event detection model aims to find clusters and each cluster denotes a real-world event containing a set of correlated social messages. Further, an incremental social event detection model detects events from continuous message blocks, which adds newly arrived messages to previous events or generates new event clusters successively.
\subsection{Overview}

When a message block $\mathrm{M}_i$ is received, the pipeline illustrated in Fig. \ref{fig:pipeline} is employed. Two different approaches are utilized to construct graphs, and node-level and graph-level contrastive learning is conducted to obtain the hybrid social message representation. Then, DRL-SinglePass clustering is used to compute the state based on the clusters of the block, and reinforcement learning is employed to learn an appropriate threshold (refer to Section \ref{section:Graph Construction}, \ref{section:Hybrid Graph Contrastive Learning} and \ref{section: DRL-SinglePass} for details). The learned threshold is applied to SinglePass clustering to cluster the current message block $\mathrm{M}_i$ and obtain predicted labels, which are compared with the ground-truth labels to derive clustering results.

\subsection{Graph Construction}
\label{section:Graph Construction}
As shown in Fig. \ref{fig:algoframe}, we adopt two methods to construct graphs to simultaneously learn the attribute information of social messages and the information interrelated between messages. The social message attribute graph emphasizes the specific characteristics and information within an individual message, while the social message relation graph focuses on learning the correlations between multiple messages.
\begin{itemize}
\item\textit{Social Message Attribute Graph.} We adopt a simple but effective graph structure for social messages. Referring to the star topology structure, we take a message as the central node, linked by its neighboring attribute nodes. Specifically, we connect the component words, location, topic, and other attributes to the central node. Then, we can obtain an attribute graph $\mathcal{G} $ containing all the information for each social message. Furthermore, when dealing with social messages from different sources, we only need to connect or remove the attribute node instead of designing a new feature acquisition approach.

\item \textit{Social Message Relation Graph.} To begin with, we integrate various attributes of messages, including words, location, topic, and other relevant characteristics, as well as users and messages themselves, as nodes in our model. We then connect messages with their respective elements, forming a heterogeneous information network graph. Then, we convert this graph into a homogeneous message graph $\mathcal{M}$, which includes only message nodes and edges connecting messages that have shared features. The transformation aims to prioritize learning correlations between messages in the homogeneous graph over retaining diverse node types in the model. As shown in Fig. \ref{fig:algoframe}, following the mapping process in KPGNN \cite{KPGNN}, we derive the homogeneous message graph $\mathcal{M}$ containing the messages $\left\{m_1, m_2, m_3, m_4, \ldots \right\}$ as nodes.

\end{itemize}

\subsection{Hybrid Graph Contrastive Learning}
\label{section:Hybrid Graph Contrastive Learning}
As shown in Fig. \ref{fig:algoframe}, we adopt two different modules, i.e., the graph-level and node-level contrastive learning modules to learn different information. 
\subsubsection{Graph Augmentation}
The beneficial augmentation types can be dataset-specific. Considering that the edge perturbation benefits social networks \cite{you2020graph}, we apply such an augmentation on two kinds of graphs. Specifically, during the edge perturbation process, we remove some edges and add more edges in the social message attribute graphs and heterogeneous social message relation graph, while avoiding the generation of new isolated points. The number of edges removed and added is equal to one-tenth of the total number of edges in the graph, and importantly, the removed edges and added edges are non-overlapping.

\subsubsection{Graph-level Contrastive Learning} 
The given social message attribute graph $\mathcal{G}$ undergoes graph data augmentations to obtain two correlated views $\hat{\mathcal{G}}_1, \hat{\mathcal{G}}_2$, as a positive pair.  
Following Equation \ref{eq:gcn}, we use the GNN-based encoder \cite{kipf2016semi} and \textit{a global attention pooling layer} as the encoder to extract \textit{graph} representations vectors $\boldsymbol{h}_1^{\mathcal{G}}, \boldsymbol{h}_2^{\mathcal{G}}$ from augmented graphs $\hat{\mathcal{G}}_1, \hat{\mathcal{G}}_2$. A two-layer perceptron, as the projection head, is applied to map $\boldsymbol{h}_1^{\mathcal{G}}, \boldsymbol{h}_2^{\mathcal{G}}$ into $\boldsymbol{z}_1^{\mathcal{G}}, \boldsymbol{z}_2^{\mathcal{G}}$ for the further contrast. 

During the training process, we consider representations of the augmented social message attribute graphs, $\boldsymbol{z}_1^{\mathcal{G}}$ and $\boldsymbol{z}_2^{\mathcal{G}}$, as positive pairs, while negative pairs are generated from the remaining augmented graphs in the message block. Then, we apply Equation \ref{eq:gcaloss2} to enforce maximizing the consistency between positive pairs $\boldsymbol{z}_1^{\mathcal{G}}, \boldsymbol{z}_2^{\mathcal{G}}$  compared with negative pairs.

\subsubsection{Node-level Contrastive Learning}
Analogically, we use the graph augmentation to get $\hat{\mathcal{M}}_1, \hat{\mathcal{M}_2}$ from the homogeneous social graph $\mathcal{M}$. Then, we follow Equation \ref{eq:gcn} to perform the GNN-based encoder, \textit{without the pooling layer}, to learn the \textit{node} representations $\boldsymbol{H}_1^{\mathcal{M}}$ and $\boldsymbol{H}_2^{\mathcal{M}}$, and the projection head is imposed to learn the ultimate representations $\boldsymbol{Z}_1^{\mathcal{M}}$ and $\boldsymbol{Z}_2^{\mathcal{M}}$.

As shown in Fig. \ref{fig:algoframe}, each node in $\mathcal{M}$ represents a social message $m$. For any node, its embeddings generated in two views form the positive sample, and the other nodes in the two views are regarded as negative samples.
Finally, the objective in Equation \ref{eq:gcaloss2} is maximized to train this module.

\subsubsection{Hybrid Embedding}
The social message attribute graph pays attention to the characteristic information of a single message, while the social heterogeneous graph focuses on learning the correlations between messages. At the testing stage, we \textit{concat} the \textit{graph} embedding of $\mathcal{G}$ and the \textit{node} embedding of $\mathcal{M}$ to derive the hybrid embedding of the social message to represent social messages comprehensively.

\subsection{Deep Reinforcement Learning Guided SinglePass}
\label{section: DRL-SinglePass}
\begin{figure}[t]
	\vskip -0in
	\begin{center}
		\centerline{\includegraphics[width=0.5\textwidth,height=0.2\textheight]{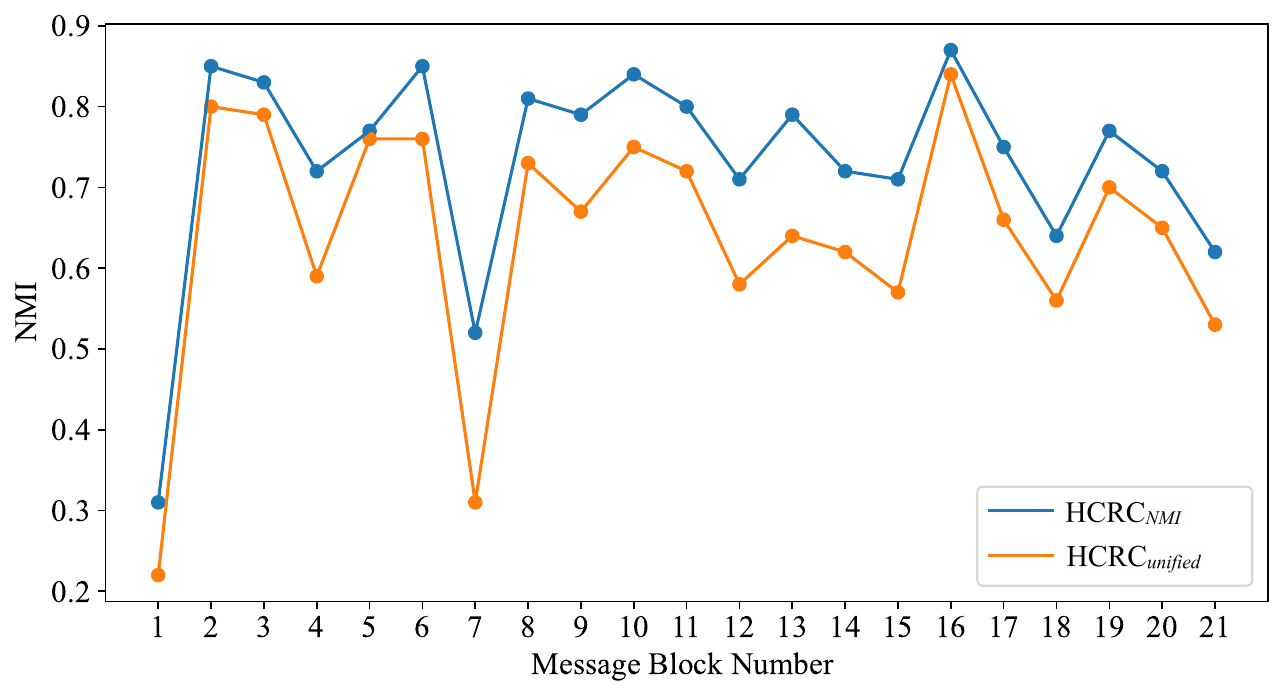}}
		\vskip -0.in
		\caption{Compare HCRC$_{NMI}$ and HCRC$_{unified}$ wth the metric of NMI. In HCRC$_{unified}$, we set the threshold of all message blocks to 0.6. The threshold of HCRC$_{NMI}$ is selected from 0.6, 0.65, 0.7, 0.75, and 0.8 based on NMI.}
		\label{fig:drltoy}
	\end{center}
\vspace{-0.5em}
\end{figure}

\begin{table*}[t] 

\small
\renewcommand\arraystretch{1.3}
\caption{The statistics of each message block in the Twitter dataset.}
\label{table: statistics}
\begin{center}
\setlength{\tabcolsep}{2mm}{
\begin{tabular}{c|ccccccccccc}
\hline
{Blocks} & {M$_{0} $} & {M$_{1} $} & {M$_{2} $} & {M$_{3} $} & {M$_{4} $} & {M$_{5} $} & {M$_{6} $} & {M$_{7} $} &{M$_{8} $} & {M$_{9} $} & {M$_{10} $} \\
\hline 
\# & 20,254 & 8,722 & 1,491 & 1,835 & 2,010 & 1,834 & 1,276 & 5,278  & 1,560 & 1,363 & 1,096  \\
\hline
\hline
{Blocks}  &{M$_{11} $}& {M$_{12} $} & {M$_{13} $} & {M$_{14} $} & {M$_{15} $} & {M$_{16} $} & {M$_{17} $} & {M$_{18} $} & {M$_{19} $} & {M$_{20} $} & {M$_{21} $}\\
\hline
\#  & 1,232& 3,237 & 1,972 & 2,956 & 2,549 &  910 & 2,676 & 1,887 & 1,399 & 893 & 2,410\\
\hline
\end{tabular}}
\end{center}
\vspace{0.5em}
\end{table*}
In real-world scenarios, it's difficult to specify hyperparameters such as $k$ for K-Means in a supervised manner since the number of topics is not available. Orthogonal to the baselines adopting K-Means in each block for clustering, we propose an improved SinglePass-based incremental clustering algorithm. Compared with K-Means, the algorithm can be completed in one pass, which is more suitable for real-world scenarios. But SinglePass is highly sensitive to the threshold value, which greatly affects the resulting clustering outcome.

We propose Deep Reinforcement Learning Guided SinglePass, dubbed DRL-SinglePass, to improve the SinglePass algorithm in streaming data. As shown in Fig. \ref{fig:drltoy}, our model can get performance boosts after adjusting the threshold from a fixed hyperparameter to a well-selected value that is continuous and trainable. Such a reinforcement learning approach enables the model to adjust the threshold adaptively instead of randomly selecting from the preset thresholds.

DRL-SinglePass regards the social data as the environment, the hyperparameter adjustment approach as an agent, and formally expresses the process as a Markov Decision Process (MDP) $\left(\mathcal{S}^{c l u}, \mathcal{A}^{c l u}, \mathcal{R}^{c l u}\right)$, where $\mathcal{S}^{c l u}$ is state space, $\mathcal{A}^{c l u}$ is action space, $\mathcal{R}^{c l u}$ is reward function. Specifically, at time step $t$, when receiving the new message block M$_t$, we define three elements as follows:

\textit{State}. The clustering result observed by the agent after each parameter adjustment episode is represented by the state, which is described as follows:
\begin{equation}
    s^{c l u}_t=\left\{\epsilon_t,\operatorname{coh}_t, \operatorname{sep}_t, DI_t, S_t\right\},
\end{equation}
which consists of the current minimum neighbor distance $\epsilon_t$, the average cohesion distance $\operatorname{coh}_t$, the average separation distance $\operatorname{sep}_t$, the Dunn Index $DI_t$ \cite{dunn1974well}, and the Silhouette coefficient $S_t$ \cite{rousseeuw1987silhouettes}. All of these serve to assess the quality of clustering and do not depend on any prior knowledge of ground-truth labels. 

\textit{Action}. 
We define the action $a^{c l u}_t$ at time step $t$ as the change of the threshold parameter of SinglePass that should be selected for the current state $s^{c l u}_t$. According to practical prior knowledge, the action space is a continuous value that is restricted in the range of $[0.6, 0.8]$.

\textit{Reward}. We apply the Calinski-Harabasz evaluation index \cite{calinski1974dendrite} as the reward function at time step $t$, of which essence is the ratio of the inter-cluster distance to the intra-cluster distance as follows:
\begin{equation}
r^{c l u}_t=\begin{array}{cl}
\frac{V^B_t}{|E|_t-1} / \frac{V^W_t}{N-|E|_t},
\end{array}
\end{equation}
where $V^W_t$ is the overall within-cluster variance and $V^B_t$ is the overall between-cluster variance. $N$ is the total number of messages. $|E|$ is the number of clusters.

\textit{Optimization}. We apply the classic policy-gradient method PPO \cite{ppo} as the updating method for the policy network.

\section{Experiment}
\subsection{Experimental Settings}
\subsubsection{Datasets}
To evaluate HCRC, we conduct experiments on two large, publicly available social media datasets, i.e., the Twitter dataset \cite{mcminn2013building} and the MAVEN dataset~\cite{Wang2020}, following KPGNN \cite{KPGNN}. After data cleaning, Twitter contains 68,841 tweets, covering 503 event classes and spreading over a period of four weeks. MAVEN is a general-domain event detection dataset, used for training and evaluating event detection systems, containing 10,242 messages and covering 164 event types. In incremental evaluation, we split Twitter into several message blocks by date. We use the first week’s messages as an initial message block M$_{0} $ and the remaining messages in Twitter to form several message blocks M$_{1} $, M$_{2} $, \ldots, and M$_{21} $ by date. Table \ref{table: statistics} shows the statistics of each message block.

\begin{table*}[t]
\small
\renewcommand\arraystretch{0.93}
\caption{Traditional incremental evaluation NMIs. The best results are in bold, and the second-best are underlined.}

\label{table: Traditional Incremental evaluation NMIs. }
\begin{center}
\setlength{\tabcolsep}{5mm}{
\begin{tabular}{c|ccccccc}
\hline
{Blocks} & {M$_{1} $} & {M$_{2} $} & {M$_{3} $} & {M$_{4} $} & {M$_{5} $} & {M$_{6} $} & {M$_{7} $} \\
\hline
Word2vec & .19±.00 & .50±.00 & .39±.00 & .34±.00 & .41±00 & .53±.00 & .25±.00\\
LDA & .11±.00 & .27±.01 & .28±.00 & .25±.00 & .26±.00 & .32±.00 & .18±.01\\
WMD & .32±.00 & .71±.00 & .67±.00 & .50±.00 & .61±.00 & .61±.00 & .46±.00\\
BERT & .36±.00 & .78±.00 & .75±.00 & .60±.00 & .72±.00 & .78±.00 & .54±.00\\
BiLSTM & .24±.00 & .50±.00 & .39±.00 & .40±.00 & .41±.00 & .50±.00 & .33±.00\\
PP-GCN & .23±.00 & .57±.02 & .55±.01 & .46±.01 & .48±.01 & .57±.01 & .37±.00\\
EventX & .36±.00 & .68±.00 & .63±.00 & .63±.00 & .59±.00 & .70±.00 & .51±.00\\
KPGNN$_{t} $ & .38±.01 & .78±.01 &  .77±.00  &  \underline{.68±.01 }&  .73±.01 & .81±.00 &  .54±.01\\
KPGNN & \underline{.39±.00} &  .79±.01 & .76±.00 & .67±.00 &  .73±.01 &  .82±.01 & \underline{.55±.01}\\
QSGNN & \textbf{.43±.01} & \underline{.81±.02} & \underline{.78±.01} & \textbf{.71±.01} & \underline{.75±.00} & \underline{.83±.01} & \textbf{.57±.01}\\
QSGNN$^*$ & .34±.02 & .73±.01 & .56±.01 & .58±.00 & .58±.02 & 71±.01 & 35±.01 \\
\hline
HCRC$_{NMI} $ & .30±.01 & \textbf{.85±.00} & \textbf{.83±.00} & \textbf{.71±.01} & \textbf{.77±.00} & \textbf{.85±.00}  & .52±.00\\
$\Delta$ & $\downarrow$ 13\% & $\uparrow$ 4\% & $\uparrow$ 5\% & $\uparrow$ 0\% & $\uparrow$ 2\% & $\uparrow$ 2\% & $\downarrow$ 5\%\\
\hline
\end{tabular}}
\begin{tabular}{c|ccccccc}
\hline
{Blocks} & {M$_{8} $} & {M$_{9} $} & {M$_{10} $} & {M$_{11} $} & {M$_{12} $} & {M$_{13} $} & {M$_{14} $} \\
\hline
Word2vec & .46±.00 & .35±.00 & .51±.00 & .37±.00 & .30±00 & .37±.00 & .36±.00\\
LDA & .37±.01 & .34±.00 & .44±.01 & .33±.01 & .22±.01 & .27±.00 & .21±.00\\
WMD & .67±.00 & .55±.00 & .61±.00 & .50±.00 & .60±.00 & .54±.00 & .66±.00\\
BERT & .79±.00 & .70±.00 & .74±.00 & .68±.00 & .59±.00 & .63±.00 & .64±.00\\
BiLSTM & .49±.00 & .43±.00 & .50±.00 & .49±.00 & .39±.00 & .46±.00 & .44±.00\\
PP-GCN & .55±.02 & .51±.02 & .55±.02 & .50±.01 & .45±.01 & .47±.01 & .44±.01\\
EventX & .71±.00 & .67±.00 & .68±.00 & .65±.00 & .61±.00 & .58±.00 & .57±.00\\
KPGNN$_{t} $ & .79±.01 & .74±.01 & .79±.00 & .73±.00 & .69±.01 & .68±.01 & .68±.01\\
KPGNN &  \underline{.80±.00} &  .74±.02 &  .80±.01 &  .74±.01 & .68±.01 &  \underline{.69±.01} &  \underline{.69±.00}\\
QSGNN  & .79±.01 & \underline{.77±.02} & \underline{.82±.02} & \underline{.75±.01} & \underline{.70±.00} & .68±.02 & .68±.01\\
QSGNN$^*$  & .61±.02 & .55±.00 & .62±.01 & .57±.01 & .45±.01 & .54±.00 & .45±.01\\
\hline
HCRC$_{NMI} $ & \textbf{.81±.00} & \textbf{.79±.00} & \textbf{.84±.00} & \textbf{.80±.00} & \textbf{.70±.01} & \textbf{.79±.00} & \textbf{.71±.01}\\
$\Delta$ &  $\uparrow$ 2\%  & $\uparrow$ 2\% & $\uparrow$ 2\% & $\uparrow$ 5\% & $\uparrow$ 0\% & $\uparrow$ 10\% &  $\uparrow$ 3\% \\
\hline
\end{tabular}
\setlength{\tabcolsep}{5mm}{
\begin{tabular}{c|ccccccc}
\hline
{Blocks} & {M$_{15} $} & {M$_{16} $} & {M$_{17} $} & {M$_{18} $} & {M$_{19} $} & {M$_{20} $} & {M$_{21} $} \\
\hline
Word2vec & .27±.00 & .49±.00 & .33±.00 & .29±.00 & .37±00 & .38±.00 & .31±.00\\
LDA & .21±.00 & .35±.01 & .19±.00 & .18±.00 & .29±.01 & .35±.00 & .19±.00\\
WMD & .51±.00 & .60±.00 & .55±.00 & .63±.00 & .54±.00 & .58±.00 & .58±.00\\
BERT & .54±.00 & .75±.00 & .63±.00 & .57±.00 & .66±.00 & .68±.00 & .59±.00\\
BiLSTM & .40±.00 & .53±.00 & .45±.00 & .44±.00 & .44±.00 & .48±.00 & .41±.00\\
PP-GCN & .39±.01 & .55±.01 & .48±.00 & .47±.01 & .51±.02 & .51±.01 & .41±.02\\
EventX & .49±.00 & .62±.00 & .58±.00 & .59±.00 & .60±.00 & .67±.00 & .53±.00\\
KPGNN$_{t} $ & .57±.01 & .78±.01 & .69±.01 &  .68±.01 & .73±.00 & \underline{.73±.00} & .59±.01\\
KPGNN &  .58±.00 &  \underline{.79±.01} &  .70±.01 & \underline{.68±.02} &  \underline{.73±.01} &  .72±.02 &  .60±.00\\
QSGNN  & \underline{.59±.01} & .78±.01 & \underline{.71±.01} & \textbf{.70±.01} & .73±.00 & \textbf{.73±.02} & \underline{.61±.01}\\
QSGNN$^*$  & .39±.01 & .55±.02 & .43±.01 & .42±.01 & .50±.00 & .52±.01 & .35±.00\\
\hline
HCRC$_{NMI} $ & \textbf{.70±.01} & \textbf{.87±.00} & \textbf{.75±.00} & .63±.01 & \textbf{.76±.01} & .72±.00 &\textbf{.62±.00}\\
$\Delta$ &  $\uparrow$ 11\%  & $\uparrow$ 8\% & $\uparrow$ 4\% & $\downarrow$ 7\% & $\uparrow$ 3\% & $\downarrow$ 1\% &  $\uparrow$ 1\% \\
\hline
\end{tabular}}
\end{center}
\end{table*}

\begin{table*}[t]
\small
\renewcommand\arraystretch{0.93}
\caption{Traditional incremental evaluation AMIs. The best results are marked in bold, and the second-best are underlined.}
\label{table: Traditional incremental evaluation AMIs.}
\begin{center}
\setlength{\tabcolsep}{5mm}{
\begin{tabular}{c|ccccccc}
\hline
{Blocks} & {M$_{1} $} & {M$_{2} $} & {M$_{3} $} & {M$_{4} $} & {M$_{5} $} & {M$_{6} $} & {M$_{7} $} \\
\hline
Word2vec & .08±.00 & .41±.00 & .31±.00 & .24±.00 & .33±00 & .40±.00 & .13±.00\\
LDA & .08±.00 & .20±.01 & .22±.01 & .17±.00 & .21±.00 & .20±.00 & .12±.01\\
WMD & .30±.00 & .69±.00 & .63±.00 & .45±.00 & .57±.00 & .57±.00 & .46±.00\\
BERT & .34±.00 & .76±.00 & .73±.00 & .55±.00 & .71±.00 & .74±.00 & .50±.00\\
BiLSTM & .12±.00 & .41±.00 & .31±.00 & .30±.00 & .33±.00 & .36±.00 & .20±.00\\
PP-GCN & .21±.00 & .55±.02 & .52±.01 & .42±.01 & .46±.01 & .52±.02 & .34±.00\\
EventX & .06±.00 & .29±.00 & .18±.00 & .19±.00 & .14±.00 & .27±.00 & .13±.00\\
KPGNN$_{t} $ & .36±.01 & .77±.01 & .75±.00 & \underline{.65±.01} & \underline{.71±.01} & .78±.00 & .50±.01\\
KPGNN & \underline{.37±.00} & .78±.01 & .74±.00 & .64±.01 & \underline{.71±.01} & .79±.01 & \underline{.51±.01}\\
QSGNN & \textbf{.41±.02} & \underline{.80±.01} & \underline{.76±.01} & \textbf{.68±.01} & \textbf{.73±.00} & \underline{.80±.01} & \textbf{.54±.00}\\
QSGNN$^*$  & .32±.01 & .70±.02 & .53±.01 & .54±.00 & .55±.01 & .65±.02 & .29±.00\\
\hline
HCRC$_{NMI} $ & .29±.01 & \textbf{.83±.00} & \textbf{.81±.01} & .64±.01 & \textbf{.73±.00} & \textbf{.81±.00} & .44±.02\\
$\Delta$ &  $\downarrow$ 12\%  & $\uparrow$ 3\% & $\uparrow$ 5\% & $\downarrow$ 4\% & $\uparrow$ 0\% & $\uparrow$ 1\% &  $\downarrow$ 10\% \\
\hline
\end{tabular}}

\begin{tabular}{c|ccccccc}
\hline
{Blocks} & {M$_{8} $} & {M$_{9} $} & {M$_{10} $} & {M$_{11} $} & {M$_{12} $} & {M$_{13} $} & {M$_{14} $} \\
\hline
Word2vec & .33±.00 & .24±.00 & .39±.00 & .26±.00 & .23±00 & .23±.00 & .26±.00\\
LDA & .24±.01 & .24±.00 & .36±.01 & .25±.01 & .16±.01 & .19±.00 & .15±.00\\
WMD & .63±.00 & .46±.00 & .57±.00 & .42±.00 & .58±.00 & .50±.00 & .64±.00\\
BERT & .75±.00 & .66±.00 & .70±.00 & .65±.00 & .56±.00 & .59±.00 & .61±.00\\
BiLSTM & .35±.00 & .32±.00 & .39±.00 & .37±.00 & .32±.00 & .31±.00 & .34±.00\\
PP-GCN & .49±.02 & .46±.02 & .51±.02 & .46±.01 & .42±.01 & .43±.01 & .41±.01\\
EventX & .21±.00 & .19±.00 & .24±.00 & .24±.00 & .16±.00 & .16±.00 & .14±.00\\
KPGNN$_{t} $ & \underline{.75±.01} & .70±.01 & .76±.01 & .70±.00 & \underline{.66±.01} & .65±.01 & .65±.01\\
KPGNN & \textbf{.76±.01} & .71±.02 & .78±.01 & .71±.01 & \underline{.66±.01} & \underline{.67±.01} & .65±.00\\
QSGNN  &\underline{.75±.01} & \textbf{.75±.02} & \underline{.80±.03} & \underline{.72±.01} & \textbf{.68±.00} & .66±.01 & \underline{.66±.01}\\
QSGNN$^*$   & .53±.00 & .48±.01 & .56±.01 & .51±.02 & .40±.00 & .48±.02 & .40±.01\\
\hline
HCRC$_{NMI} $ &\underline{.75±.01} & \underline{.72±.01} & \textbf{.82±.00} & \textbf{.76±.00} & .62±.02 & \textbf{.76±.00} & \textbf{.67±.01}\\
$\Delta$ &  $\downarrow$ 1\%  & $\downarrow$ 3\% & $\uparrow$ 2\% & $\uparrow$ 4\%  & $\downarrow$ 6\% & $\uparrow$ 9\% &  $\uparrow$ 1\% \\
\hline
\end{tabular}

\begin{tabular}{c|ccccccc}
\hline
{Blocks} & {M$_{15} $} & {M$_{16} $} & {M$_{17} $} & {M$_{18} $} & {M$_{19} $} & {M$_{20} $} & {M$_{21} $} \\
\hline
Word2vec & .15±.00 & .36±.00 & .24±.00 & .21±.00 & .28±00 & .24±.00 & .21±.00\\
LDA & .13±.00 & .27±.01 & .13±.00 & .12±.00 & .22±.01 & .23±.00 & .13±.00\\
WMD & .47±.00 & .59±.00 & .57±.00 & .60±.00 & .49±.00 & .55±.00 & .52±.00\\
BERT & .50±.00 & .72±.00 & .60±.00 & .53±.00 & .63±.00 & .62±.00 & .57±.00\\
BiLSTM & .26±.00 & .41±.00 & .35±.00 & .35±.00 & .35±.00 & .34±.00 & .31±.00\\
PP-GCN & .35±.01 & .52±.01 & .45±.00 & .45±.01 & .48±.02 & .45±.02 & .38±.02\\
EventX & .07±.00 & .19±.00 & .18±.00 & .16±.00 & .16±.00 & .18±.00 & .10±.00\\
KPGNN$_{t} $ & .53±.01 & .75±.01 & .67±.01 & .66±.01 & .70±.00 & .68±.00 & \underline{.57±.01}\\
KPGNN & .54±.00 & \underline{.77±.01} & .68±.01 & \underline{.66±.02} & \underline{.71±.01} & \underline{.68±.02} & .57±.00\\
QSGNN   &\underline{.55±.01} & .76±.02 & \underline{.69±.01} & \textbf{.68±.01} & .70±.01 & \textbf{.69±.02} & \textbf{.58±.00}\\
QSGNN$^*$  & .33±.01 & 49±.00 & .40±.00 & .36±.01 & .45±.02 & .42±.02 & .31±.01 \\
\hline
HCRC$_{NMI} $ & \textbf{.66±.01} & \textbf{.86±.00} & \textbf{.72±.00} &.50±.03 &\textbf{.72±.01} & .61±.00 & .55±.01 \\
$\Delta$ &  $\uparrow$ 11\%  & $\uparrow$ 9\% & $\uparrow$ 3\% & $\downarrow$ 18\% & $\uparrow$ 1\% & $\downarrow$ 8\% & $\downarrow$ 3\% \\
\hline
\end{tabular}
\end{center}
\end{table*}

\begin{table*}[t]
\small
\renewcommand\arraystretch{0.93}

\caption{Traditional incremental evaluation ARIs. The best results are marked in bold, and the second-best are underlined.}
\label{table: Traditional incremental evaluation ARIs.}
\begin{center}
\setlength{\tabcolsep}{5mm}{
\begin{tabular}{c|ccccccc}
\hline
{Blocks} & {M$_{1} $} & {M$_{2} $} & {M$_{3} $} & {M$_{4} $} & {M$_{5} $} & {M$_{6} $} & {M$_{7} $} \\
\hline
Word2vec & .01±.00 & .49±.00 & .16±.00 & .07±.00 & .17±00 & .25±.00 & .02±.00\\
LDA & .00±.00 & .08±.00 & .02±.01 & .07±.00 & .06±.00 & .07±.01 & .00±.00\\
WMD & .04±.00 & .48±.00 & .28±.00 & .11±.00 & .26±.00 & .16±.00 & .08±.00\\
BERT & .03±.00 & .64±.00 & .43±.00 & .19±.00 & .44±.00 & .44±.00 & .07±.00\\
BiLSTM & .03±.00 & .49±.00 & .17±.00 & .11±.00 & .19±.00 & .18±.00 & .12±.00\\
PP-GCN & .05±.00 & .67±.03 & .47±.01 & .24±.01 & .34±.00 & .55±.03 & .11±.02\\
EventX & .01±.00 & .45±.00 & .09±.00 & .07±.00 & .04±.00 & .14±.00 & .02±.00\\
KPGNN$_{t} $ & .06±.01 & .76±.01 & \underline{.60±.02} & \underline{.30±.01} & \underline{.48±.01} & .67±.05 & .11±.01\\
KPGNN & .07±.01 & \underline{.76±.02} & .58±.01 & .29±.01 & .47±.03 & \underline{.72±.03} & .12±.00\\
QSGNN & - & - &- &- &- &- &- \\
QSGNN$^*$  & \underline{.13±.02} & .74±.00 & .36±.01 & .28±.01 & .32±.00 & .45±.01 & \underline{.13±.01}\\
\hline
HCRC$_{NMI} $ & \textbf{.18±.05} & \textbf{.82±.00} & \textbf{.70±.01} & \textbf{.41±.01} & \textbf{.60±.01} & \textbf{.81±.01} & \textbf{.19±.02}\\
$\Delta$ &  $\uparrow$ 5\%  & $\uparrow$ 6\% & $\uparrow$ 10\% & $\uparrow$ 11\% & $\uparrow$ 12\% & $\uparrow$ 9\% &  $\uparrow$ 6\% \\
\hline
\end{tabular}}

\begin{tabular}{c|ccccccc}
\hline
{Blocks} & {M$_{8} $} & {M$_{9} $} & {M$_{10} $} & {M$_{11} $} & {M$_{12} $} & {M$_{13} $} & {M$_{14} $} \\
\hline
Word2vec & .17±.00 & .08±.00 & .23±.00 & .09±.00 & .09±00 & .06±.00 & .10±.00\\
LDA & .03±.01 & .03±.01 & .09±.02 & .03±.01 & .02±.00 & .00±.00 & .02±.00\\
WMD & .22±.00 & .12±.00 & .20±.00 & .12±.00 & .27±.00 & .13±.00 & .33±.00\\
BERT & .50±.00 & .33±.00 & .44±.00 & .27±.00 & .31±.00 & .14±.00 & .30±.00\\
BiLSTM & .17±.00 & .13±.00 & .30±.00 & .16±.00 & .14±.00 & .10±.00 & .17±.00\\
PP-GCN & .43±.04 & .31±.02 & .50±.07 & .38±.02 & .34±.03 & .19±.01 & .29±.01\\
EventX & .09±.00 & .07±.00 & .13±.00 & .16±.00 & .07±.00 & .04±.00 & .10±.00\\
KPGNN$_{t} $ & \underline{.59±.02} & .45±.02 & .64±.01 & .48±.01 & \textbf{.50±.03} & .28±.01 & \underline{.43±.02}\\
KPGNN & \textbf{.60±.01} & \underline{.46±.02} & \underline{.70±.06} & \underline{.49±.03} & \underline{.48±.01} & \underline{.29±.03} & .42±.02\\
QSGNN  & - & - &- &- &- &- &- \\
QSGNN$^*$  & .31±.00 & .26±.02 & .40±.01 & .26±.00 & .22±.00 & .25±.01 & .22±.01 \\
\hline
HCRC$_{NMI} $ & .58±.01 & \textbf{.55±.02} & \textbf{.81±.00} & \textbf{.78±.00} & .44±.07 & \textbf{.72±.00} & \textbf{.54±.03} \\
$\Delta$& $\downarrow$ 2\% & $\uparrow$ 9\% & $\uparrow$ 11\% & $\uparrow$ 29\% & $\downarrow$ 6\% & $\uparrow$ 43\% &  $\uparrow$ 11\% \\
\hline
\end{tabular}

\begin{tabular}{c|ccccccc}
\hline
{Blocks} & {M$_{15} $} & {M$_{16} $} & {M$_{17} $} & {M$_{18} $} & {M$_{19} $} & {M$_{20} $} & {M$_{21} $} \\
\hline
Word2vec & .03±.00 & .19±.00 & .10±.00 & .07±.00 & .14±00 & .10±.00 & .06±.00\\
LDA & .00±.00 & .11±.01 & .02±.00 & .02±.00 & .03±.00 & .02±.01 & .00±.01\\
WMD & .16±.00 & .32±.00 & .26±.00 & .35±.00 & .12±.00 & .19±.00 & .19±.00\\
BERT & .10±.00 & .41±.00 & .24±.00 & .24±.00 & .32±.00 & .33±.00 & .18±.00\\
BiLSTM & .08±.00 & .27±.00 & .22±.00 & .19±.00 & .16±.00 & .20±.00 & .16±.00\\
PP-GCN & .15±.00 & .51±.03 & .35±.03 & .39±.03 & .41±.02 & .41±.01 & .20±.03\\
EventX & .01±.00 & .08±.00 & .12±.00 & .08±.00 & .07±.00 & .11±.00 & .01±.00\\
KPGNN$_{t} $ & .16±.02 & .62±.03 & .41±.03 & \underline{.46±.02} & .50±.01 & \underline{.51±.01} & .01±.00\\
KPGNN & \underline{.17±.00} & \underline{.66±.05} & \underline{.43±.05} & \textbf{.47±.04} & \underline{.51±.03} & \textbf{.51±.04} & \underline{.20±.01}\\
QSGNN & - & - &- &- &- &- &- \\
QSGNN$^*$ & .13±.01 & .34±.01 & .22±.02 & .22±.00 & .28±.01 & .24±.00 & .13±.01 \\
\hline
HCRC$_{NMI} $ & \textbf{.70±.05} & \textbf{.87±.00} & \textbf{.70±.04} & .32±.06 & \textbf{.58±.03} & .40±.00 & \textbf{.36±.00}\\
$\Delta$ &  $\uparrow$ 53\% & $\uparrow$ 21\% & $\uparrow$ 27\% & $\downarrow$ 15\% & $\uparrow$ 7\% & $\downarrow$ 11\% &  $\uparrow$ 16\% \\
\hline
\end{tabular}
\end{center}
\end{table*}

\begin{table*}[t]
\small
\renewcommand\arraystretch{0.94}
\caption{Semi-supervised and solid incremental evaluation NMIs. The best results are marked in bold.}
\label{table: Semi-supervised and Solid Incremental evaluation NMIs.}
\begin{center}
\setlength{\tabcolsep}{2mm}{
\begin{tabular}{c|ccccccc}
\hline
{Blocks}  & {M$_{1} $}  & {M$_{2} $}  & {M$_{3} $} & {M$_{4} $} & {M$_{5} $} & {M$_{6} $} &  {M$_{7} $}\\
\hline
KPGNN$_{10\%} $ & \textbf{.27±.01} & .68±.01 & .60±.01 &.57±.01 & .54±.02 & .70±.02 & \textbf{.37±.01} \\
QSGNN$_{10\%}$ & .25±.01 & .75±.00 & .65±.01 & .59±.02 & .60±.01 & .65±.01 & .34±.02 \\
HCRC$_{10\%} $ & .24±.03 & \textbf{.82±.02} & \textbf{.79±.01} & \textbf{.70±.01}  & \textbf{.76±.00}  & \textbf{.81±.02} & .34±.01 \\
$\Delta$ & $\downarrow$ 2\% & $\uparrow$ 7\% & $\uparrow$ 14\% & $\uparrow$ 11\%  & $\uparrow$ 16\% & $\uparrow$ 11\%  &  $\downarrow$ 3\% \\
\hline
KPGNN$_{random} $  & .27±.01 & .71±.01 & .64±.02 & .59±.01  & .61±.03 & .71±.03 & .42±.02 \\
QSGNN$_{random}$  & \textbf{.31±.02} & .77±.01 & .65±.01 & .50±.00 & .60±.02 & .75±.00 & .39±.01 \\
HCRC$_{random} $  & .27±.01 & .82±.00 & .79±.03 &.63±.05 & .70±.01 & .80±.00 & .46±.05 \\
HCRC & .27±.00 & \textbf{.83±.00} & \textbf{.81±.01} & \textbf{.67±.03} & \textbf{.74±.01} & \textbf{.83±.01} & \textbf{.50±.01}\\
$\Delta$ & $\downarrow$ 4\% & $\uparrow$ 6\% & $\uparrow$ 16\% & $\uparrow$ 8\%  & $\uparrow$ 13\% & $\uparrow$ 8\%  &  $\uparrow$ 8\%  \\
\hline
\end{tabular}}

\setlength{\tabcolsep}{2mm}{
\begin{tabular}{c|ccccccc}
\hline
{Blocks}  & {M$_{8} $}  & {M$_{9} $} & {M$_{10} $} & {M$_{11} $} & {M$_{12} $} & {M$_{13} $} & {M$_{14} $}\\
\hline
KPGNN$_{10\%} $ & .69±.02 & .55±.03 & .68±.03 & .61±.02 & .47±.02 & .56±.05 & .41±.03 \\
QSGNN$_{10\%}$ & .60±.03 & .52±.01 & .63±.00 & .57±.00 & .46±.02 & .56±.01 & .44±.02\\
HCRC$_{10\%} $ & \textbf{.79±.00} & \textbf{.75±.02} & \textbf{.74±.03} & \textbf{.78±.02} & \textbf{.68±.02} & \textbf{.76±.04} & \textbf{.65±.02} \\
$\Delta$ & $\uparrow$ 10\%  & $\uparrow$ 20\% & $\uparrow$ 6\% & $\uparrow$ 17\%  & $\uparrow$ 21\% & $\uparrow$ 20\%  &  $\uparrow$ 21\%  \\
\hline
KPGNN$_{random} $  & .72±.02 & .62±.03 &.69±.02 & .64±.01 & .51±.03 & .58±.01 & .50±.04 \\
QSGNN$_{random}$ & .71±.00 & .68±.02 & .65±.01 & .52±.01 & .63±.00 & .49±.01 & .50±.00 \\
HCRC$_{random} $  & .75±.01 & .70±.01 & .78±.01 & .67±.01 & .65±.01 & .67±.01 & \textbf{.68±.01} \\
HCRC & \textbf{.78±.01} & \textbf{.76±.00} & \textbf{.80±.01} & \textbf{.70±.02} & \textbf{.69±.00} & \textbf{.69±.04} & \textbf{.68±.01}\\
$\Delta$ & $\uparrow$ 6\%  & $\uparrow$ 8\% & $\uparrow$ 11\% & $\uparrow$ 6\%  & $\uparrow$ 6\% & $\uparrow$ 11\%  &  $\uparrow$ 18\%  \\
\hline
\end{tabular}}

\setlength{\tabcolsep}{2mm}{
\begin{tabular}{c|ccccccc}
\hline
{Blocks}  & {M$_{15} $} & {M$_{16} $} & {M$_{17} $}  & {M$_{18} $} & {M$_{19} $} & {M$_{20} $} & {M$_{21} $}\\
\hline
KPGNN$_{10\%} $ & .27±.04 & .71±.02 & .48±.03 & .36±.03 & .49±.02 & .53±.04 & .37±.01 \\
QSGNN$_{10\%}$ & .42±.00 & .63±.01 & .43±.00 & .51±.03 & .52±.00 & .53±.01 & .37±.01\\
HCRC$_{10\%} $ & \textbf{.59±.00} & \textbf{.85±.01} & \textbf{.65±.07} & \textbf{.62±.01} & \textbf{.75±.03} & \textbf{.64±.05} & \textbf{.57±.04} \\
$\Delta$ & $\uparrow$ 17\%  & $\uparrow$ 14\% & $\uparrow$ 17\% & $\uparrow$ 11\%  & $\uparrow$ 23\% & $\uparrow$ 11\%  &  $\uparrow$ 20\%  \\
\hline
KPGNN$_{random} $  & .41±.03 & .67±.03 & .58±.03 & .48±.06 & .57±.02 & .63±.02 & .45±.04 \\
QSGNN$_{random}$  & .51±.01 & .58±.00 & .56±.01 & .45±.01 & .58±.00 & .64±.02 & .42±.00 \\
HCRC$_{random} $  & .58±.01 & .80±.02  & .67±.01 & .60±.01 & .69±.02 & .69±.01 & .55±.03 \\
HCRC & \textbf{.68±.02} & \textbf{.86±.02} & \textbf{.71±.02} & \textbf{.61±.01} & \textbf{.74±.01} & \textbf{.69±.02} & \textbf{.56±.01} \\
$\Delta$ & $\uparrow$ 17\%  & $\uparrow$ 19\% & $\uparrow$ 13\% & $\uparrow$ 13\%  & $\uparrow$ 16\% & $\uparrow$ 5\%  &  $\uparrow$ 11\%  \\
\hline
\end{tabular}}
\end{center}
\end{table*}

\begin{table*}[t]
\small
\renewcommand\arraystretch{0.94}
\caption{Semi-supervised and solid incremental evaluation AMIs. The best results are marked in bold.}
\label{table: Semi-supervised and Solid Incremental evaluation AMIs.}
\begin{center}
\setlength{\tabcolsep}{2mm}{
\begin{tabular}{c|ccccccc}
\hline
{Blocks}  & {M$_{1} $} & {M$_{2} $} & {M$_{3} $}  & {M$_{4} $} & {M$_{5} $} & {M$_{6} $} & {M$_{7} $}\\
\hline
KPGNN$_{10\%} $ & \textbf{.27±.01} & .66±.00 & .57±.02 & .54±.01 & .53±.01 & .68±.02 & \textbf{.36±.01} \\
QSGNN$_{10\%}$  & .24±.00 & .73±.00 & .64±.01 & .56±.02 & .58±.01 & .61±.00 & .31±.01\\
HCRC$_{10\%} $ & .22±.02 & \textbf{.77±.03} & \textbf{.77±.02} & \textbf{.64±.01} & \textbf{.73±.01} & \textbf{.73±.03} & .31±.02 \\
$\Delta$ & $\downarrow$ 5\%  & $\uparrow$ 4\% & $\uparrow$ 13\% & $\uparrow$ 8\%  & $\uparrow$ 15\% & $\uparrow$ 5\%  &  $\downarrow$ 5\% \\
\hline
KPGNN$_{random} $  & .24±.01 & .67±.02 & .58±.03 & .54±.03 & .57±.04 & .64±.03 & .36±.02 \\
QSGNN$_{random}$  & \textbf{.28±.02} & .75±.00 & .50±.01 & .57±.00 & .64±.00 & .70±.02 & .28±.01 \\
HCRC$_{random} $  & .14±.01 &.77±.00  & .73±.03 & .56±.01 & .60±.02 & .73±.01 & .37±.03 \\
HCRC & .19±.00 &  \textbf{.81±.01} & \textbf{.77±.01} & \textbf{.61±.02} & \textbf{.67±.01} & \textbf{.77±.04} & \textbf{.43±.01}  \\
$\Delta$ & $\downarrow$ 9\%  & $\uparrow$ 6\% & $\uparrow$ 19\% & $\uparrow$ 4\%  & $\uparrow$ 3\% & $\uparrow$ 7\%  &  $\uparrow$ 7\%  \\
\hline
\end{tabular}}

\setlength{\tabcolsep}{2mm}{
\begin{tabular}{c|ccccccc}
\hline
{Blocks}  & {M$_{8} $}  & {M$_{9} $} & {M$_{10} $} & {M$_{11} $} & {M$_{12} $} & {M$_{13} $} & {M$_{14} $}\\
\hline
KPGNN$_{10\%} $ & .66±.03 & .49±.02 & .65±.03 & .59±.02 & .46±.02 & .57±.02 & .40±.01 \\
QSGNN$_{10\%}$ & .54±.02 & .47±.00 & .58±.00 & .54±.01 & .44±.01 & .53±.00 & .42±.01\\
HCRC$_{10\%} $ & \textbf{.73±.02} & \textbf{.71±.02} & \textbf{.71±.03} & \textbf{.73±.03} & \textbf{.62±.01} & \textbf{.74±.04} & \textbf{.62±.02} \\
$\Delta$ & $\uparrow$ 7\% & $\uparrow$ 22\% & $\uparrow$ 6\% & $\uparrow$ 14\%  & $\uparrow$ 16\% & $\uparrow$ 17\%  &  $\uparrow$ 20\%  \\
\hline
KPGNN$_{random} $  & .65±.02 & .56±.03 &.60±.03  & .56±.04 & .45±.03 & .51±.02 & .44±.04 \\
QSGNN$_{random}$  & .63±.00 & .55±.00 & .64±.01 & .59±.01 & .44±.02 & .58±.02 & .42±.00 \\
HCRC$_{random} $  & .60±.00 & .63±.00 & .71±.03 & .54±.01 & .45±.01 & .57±.01 & .60±.03 \\
HCRC & \textbf{.71±.01} & \textbf{.69±.00} & \textbf{.75±.01} & \textbf{.60±.03} & \textbf{.59±.01} & \textbf{.66±.04} & \textbf{.61±.01}  \\
$\Delta$ & $\uparrow$ 6\% & $\uparrow$ 13\% & $\uparrow$ 11\% & $\uparrow$ 1\%  & $\uparrow$ 14\% & $\uparrow$ 8\%  &  $\uparrow$ 17\%  \\
\hline
\end{tabular}}

\setlength{\tabcolsep}{2mm}{
\begin{tabular}{c|ccccccc}
\hline
{Blocks}  & {M$_{15} $} & {M$_{16} $} & {M$_{17} $}  & {M$_{18} $} & {M$_{19} $} & {M$_{20} $} & {M$_{21} $}\\
\hline
KPGNN$_{10\%} $ & .28±.02 & .67±.01 & .48±.03 & .35±.03 & .46±.02 & .51±.02 & .35±.01 \\
QSGNN$_{10\%}$  & .40±.01 & .58±.01 & .42±.00 & .38±.02 & .49±.01 & .47±.01 & .33±.00\\
HCRC$_{10\%} $ & \textbf{.56±.00} & \textbf{.82±.01} & \textbf{.64±.07} & \textbf{.50±.03} & \textbf{.71±.02} & \textbf{.56±.04} & \textbf{.52±.04} \\
$\Delta$ & $\uparrow$ 16\% & $\uparrow$ 15\% & $\uparrow$ 16\% & $\uparrow$ 12\%  & $\uparrow$ 22\% & $\uparrow$ 5\%  &  $\uparrow$ 17\%  \\
\hline
KPGNN$_{random} $  & .34±.02 & .60±.05 & .54±.04 & .42±.05 & .51±.03 & .51±.04 & .39±.04 \\
QSGNN$_{random}$  & .43±.01 & .56±.00 & .51±.00 & .39±.02 & .53±.01 & .54±.00 & .37±.01 \\
HCRC$_{random} $  & .43±.01 & .75±.03 & .66±.01 & \textbf{.52±.00} & .59±.01 & \textbf{.58±.01} & \textbf{.51±.02} \\
HCRC  & \textbf{.61±.03} & \textbf{.83±.02} & \textbf{.66±.05} & .43±.06 & \textbf{.70±.03} & .47±.08 & .36±.04\\
$\Delta$ & $\uparrow$ 18\% & $\uparrow$ 23\% & $\uparrow$ 12\% & $\uparrow$ 10\%  & $\uparrow$ 17\% & $\uparrow$ 4\%  &  $\uparrow$ 12\%  \\
\hline
\end{tabular}}
\end{center}
\end{table*}

\begin{table*}
\small
\renewcommand\arraystretch{0.8}
\setlength{\belowcaptionskip}{-0.15cm}
\caption{Semi-supervised and solid incremental evaluation ARIs. The best results are marked in bold.}
\label{table: Semi-supervised and Solid Incremental evaluation ARIs.}
\begin{center}

\setlength{\tabcolsep}{2mm}{
\begin{tabular}{c|ccccccc}
\hline
{Blocks}  & {M$_{1} $} & {M$_{2} $} & {M$_{3} $}  & {M$_{4} $} & {M$_{5} $}  & {M$_{6} $} & {M$_{7} $}\\
\hline
KPGNN$_{10\%} $ & \textbf{.12±.01} & .66±.00 & .42±.00 & .27±.02 & .35±.02 & .63±.04 & \textbf{.19±.02} \\
QSGNN$_{10\%}$  & .08±.02 & .64±.02 & .45±.00 & .30±.02 & .39±.01 & .54±.01 & .15±.00\\
HCRC$_{10\%} $ & .05±.01 & \textbf{.78±.02} & \textbf{.69±.09} & \textbf{.34±.00} &\textbf{.55±.00} & \textbf{.75±.02} & .10±.04 \\
$\Delta$ & $\downarrow$ 7\% & $\uparrow$ 12\% & $\uparrow$ 24\% & $\uparrow$ 4\%  & $\uparrow$ 16\% & $\uparrow$ 12\%  &  $\downarrow$ 9\% \\
\hline
KPGNN$_{random} $  & .02±.01 & .63±.05 & .31±.08 & .20±.06 & .30±.08 & .40±.03 & .04±.00 \\
QSGNN$_{random}$  & \textbf{.04±.00} & .68±.02 & .33±.01 & .32±.01 & .33±.03 & .41±.01 & \textbf{.13±.00} \\
HCRC$_{random} $  & .02±.00 & .78±.01 & .57±.02 & .35±.02 & .37±.02 & .74±.01 &.07±.02 \\
HCRC & .01±.00 & \textbf{.79±.00} & \textbf{.59±.04} & \textbf{.42±.05} & \textbf{.47±.02} & \textbf{.78±.04} & .11±.01\\
$\Delta$ & $\downarrow$ 3\% & $\uparrow$ 11\% & $\uparrow$ 26\% & $\uparrow$ 10\%  & $\uparrow$ 14\% & $\uparrow$ 37\%  &  $\downarrow$ 2\% \\
\hline
\end{tabular}}

\setlength{\tabcolsep}{2mm}{
\begin{tabular}{c|ccccccc}
\hline
{Blocks}  & {M$_{8} $}  & {M$_{9} $}  & {M$_{10} $} & {M$_{11} $} & {M$_{12} $} & {M$_{13} $} & {M$_{14} $}\\
\hline
KPGNN$_{10\%} $ & .53±.01 & .30±.01 & \textbf{.57±.02} & .45±.02 & .29±.02 & .34±.04 & .29±.00 \\
QSGNN$_{10\%}$  & .34±.01 & .33±.00 & .49±.02 & .44±.00 & .26±.02 & .33±.01 & .29±.01\\
HCRC$_{10\%} $ & \textbf{.53±.05} & \textbf{.53±.01} & .52±.01 & \textbf{.70±.11} & \textbf{.42±.04} & \textbf{.67±.01} & \textbf{.41±.04} \\
$\Delta$ & $\uparrow$ 0\% & $\uparrow$ 20\% & $\downarrow$ 5\% & $\uparrow$ 25\%  & $\uparrow$ 13\% & $\uparrow$ 33\%  &  $\uparrow$ 12\% \\
\hline
KPGNN$_{random} $  & .45±.02 & .28±.04 & .27±.03 & .25±.08 & .23±.02 & .16±.09 & .17±.02 \\
QSGNN$_{random}$  & .38±.01 & .32±.00 & .43±.02 & .32±.01 & .21±.03 & .49±.00 &.20±.01 \\
HCRC$_{random} $  & .37±.02 & .38±.02 & .51±.07 & .22±.02 & .17±.02 & .29±.01 & .39±.05 \\
HCRC & \textbf{.48±.01} & \textbf{.42±.01} & \textbf{.64±.02} & \textbf{.37±.07} & \textbf{.31±.02} & \textbf{.56±.06} & \textbf{.41±.00}\\
$\Delta$ & $\uparrow$ 3\%  & $\uparrow$ 10\% & $\uparrow$ 21\%  & $\uparrow$ 5\%  & $\uparrow$ 8\% & $\uparrow$ 7\%  &  $\uparrow$ 21\% \\
\hline
\end{tabular}}

\setlength{\tabcolsep}{2mm}{
\begin{tabular}{c|ccccccc}
\hline
{Blocks}  & {M$_{15} $}  & {M$_{16} $} & {M$_{17} $}  & {M$_{18} $} & {M$_{19} $} & {M$_{20} $} & {M$_{21} $}\\
\hline
KPGNN$_{10\%} $ & .11±.04 & .57±.01 & .36±.03 & .15±.04 & .21±.04 & \textbf{.36±.03} & .10±.01 \\
QSGNN$_{10\%}$  & .24±.01 & .43±.01 & .29±.02 & .23±.00 & .29±.01 & .33±.01 & .18±.00 \\
HCRC$_{10\%} $ & \textbf{.69±.00} & \textbf{.78±.03} & \textbf{.58±.13} & \textbf{.26±.04} & \textbf{.52±.03} & .28±.04 & \textbf{.42±.03} \\
$\Delta$ & $\uparrow$ 45\%  & $\uparrow$ 21\% & $\uparrow$ 22\%  & $\uparrow$ 3\%  & $\uparrow$ 23\% & $\downarrow$ 8\% &  $\uparrow$ 24\% \\
\hline
KPGNN$_{random} $  & .06±.01 & .30±.02 & .26±.02 & .18±.03 & .27±.08 & .27±.07 & .09±.02 \\
QSGNN$_{random}$  & .23±.01 & .45±.01 & .36±.00 & \textbf{.24±.01} & .32±.01 & \textbf{.38±.02} &.20±.01 \\
HCRC$_{random} $  & .26±.01 & .65±.05 & .45±.02 & .19±.01 & .39±.01 & .22±.02 & \textbf{.28±.04} \\
HCRC & \textbf{.59±.04} & \textbf{.81±.03} & \textbf{.53±.08} & .20±.08 & \textbf{.50±.05} & .21±.07 & .07±.02\\
$\Delta$ & $\uparrow$ 36\%  & $\uparrow$ 36\% & $\uparrow$ 17\%  & $\downarrow$ 4\%  & $\uparrow$ 18\% & $\downarrow$ 17\% &  $\uparrow$ 8\% \\
\hline
\end{tabular}}
\end{center}
\vspace{-2.5em}
\end{table*}

\subsubsection{Baselines}
We compare the proposed HCRC with eleven baselines, including Word2vec \cite{mikolov2013efficient}, LDA \cite{blei2003latent}, WMD \cite{kusner2015word}, BERT \cite{devlin2018bert}, BiLSTM \cite{graves2005framewise}, PP-GCN \cite{peng2019fine}, EventX \cite{liu2020story}, KPGNN \cite{KPGNN}, KPGNN$_{t}$ \cite{KPGNN}, QSGNN \cite{QSGNN} and QSGNN$^*$. \textit{Word2vec} converts all words in a message to vectors, calculates their average as the representation of the message.
\textit{LDA} is a generative model that utilizes latent topics and word distributions to obtain representations of messages.
\textit{WMD} measures the similarity between two messages by calculating the minimum distance between word embeddings in one message and the word embeddings in another message.
\textit{BERT} utilizes large-scale unlabeled corpora for training to obtain word embeddings of the words in a message, and takes the average of these word embeddings as the representation of the message.
\textit{BiLSTM} learns the bidirectional dependency between a word and other words, capturing the contextual information in a message to obtain the representation of the message.
\textit{PP-GCN} is a fine-grained social event detection method based on GCN.
\textit{EventX} is a model that performs online event detection on streaming text data.
\textit{KPGNN} is an incremental social event detection method via heterogeneous graph neural network.
\textit{KPGNN$_{t}$} removes the global-local pair loss term from the loss function of KPGNN and only utilizes the triplet loss term.
\textit{QSGNN} is a social event detection method based on quality-aware self-improving graph neural network.
\textit{QSGNN$^*$} are implemented based on the official code provided by QSGNN \cite{QSGNN}.


We further compare KPGNN, QSGNN and HCRC in different settings. Specifically, HCRC$_{NMI} $'s threshold for every message block is determined by NMI \cite{estevez2009normalized}; HCRC$_{10\%} $'s threshold, KPGNN$_{10\%} $'s $k$ and QSGNN$_{10\%} $'s $k$ for every message block are determined by using 10\% of the ground-truth label; HCRC$_{random} $'s threshold, KPGNN$_{random} $'s $k$ and QSGNN$_{random} $'s $k$ for every message block are randomly determined. HCRC's threshold is learned by the proposed DRL-SinglePass. $k$ and the threshold are the hyperparameters for K-Means in KPGNN and QSGNN and SinglePass in HCRC, respectively.

\subsubsection{Implementation Details}
The number of units in each layer of the GCN is set to 256. Moreover, the learning rates for graph-level and node-level contrastive learning are set to 6e-7 and 1e-5, respectively. Additionally, we set the moving average decay for the teacher network to 0.9. In HCRC$_{NMI}$, we employ the SinglePass clustering method with varying thresholds to obtain multiple clustering results from the pre-trained message representations. The final clustering result is obtained by selecting the one with the highest NMI score. For HCRC$_{10\%}$, we use SinglePass clustering with varying thresholds on 10\% labeled data to determine the optimal threshold for the entire message block, and then perform clustering once again to generate the final result. In DRL-SinglePass, we set the learning range to be between 0.6 and 0.8 based on the experimental results obtained from HCRC$_{NMI}$, where the majority of message blocks achieved optimal results. In DRL-SinglePass, we adopt a pre-clustering approach wherein one-tenth of the tweets within each message block are initially grouped. Subsequently, we leverage reinforcement learning to determine an optimal threshold for the entire message block, utilizing the unsupervised clustering evaluation results as a basis for learning. To ensure a fair comparison, we randomly run the experiments ten times and report the average results with standard deviations. Our implementation is available at \url{https://github.com/guoyy49/HCRC}.

\subsubsection{Evaluation Metrics}
To evaluate HCRC and baselines, we use normalized mutual information (NMI) \cite{estevez2009normalized}, adjusted mutual information(AMI) \citep{xuan2010information} and adjusted rand index (ARI) \cite{xuan2010information} to measure the similarities between the detected message clusters and the ground-truth clusters. NMI is one of the vital metrics for social event detection, which measures the similarity of clustering results ranging from 0 to 1. A higher NMI value signifies a stronger alignment between the detected message clusters and the ground-truth clusters, indicating a more successful clustering process. Conversely, a lower NMI value suggests a greater divergence between the clustering results and the true cluster assignments, indicating a potential mismatch or inconsistency in the clustering outcomes \citep{liu2020story,peng2019fine}.  AMI penalizes random assignment of cluster labels to ensure that the quality of clustering is not overestimated due to randomness. The typical range of AMI values is between -1 and 1, where 0 signifies similarity between the clustering results and true labels that is equivalent to random assignment. A score of 1 indicates a perfect match, meaning that the clustering results are identical to the true labels. Values below 0 imply that the similarity between the clustering results and true labels is worse than random assignment, possibly suggesting the negative correlation \citep{xuan2010information}. The Rand Index (RI) measures the proportion of ``correct decision-making" in clustering analysis. It compares the similarity between pairs of samples in their true labels and the clustering results. On the other hand, ARI is a normalized version of the Rand Index that ranges from -1 to 1. A higher ARI value indicates a better clustering effect, where 1 represents a perfect clustering result and 0 indicates a random distribution. Conversely, a negative ARI value suggests that the clustering result is worse than random chance \citep{xuan2010information}. Note that the results of all baselines in the offline evaluation and traditional incremental clustering refer to KPGNN \cite{KPGNN} and QSGNN\cite{QSGNN}.


\begin{table*}[t]
\small
\renewcommand\arraystretch{1.0}
\setlength{\belowcaptionskip}{-0.15cm} 
\caption{Thresholds learned by the proposed DRL-SinglePass.}
\label{table: thresholds}
\begin{center}
\setlength{\tabcolsep}{2mm}{
\begin{tabular}{c|ccccccccccc}
\hline
{Blocks} & {M$_{0} $} & {M$_{1} $} & {M$_{2} $} & {M$_{3} $} & {M$_{4} $} & {M$_{5} $} & {M$_{6} $} & {M$_{7} $} &{M$_{8} $} & {M$_{9} $} & {M$_{10} $} \\
\hline 
Threshold & - & 0.68 & 0.67 & 0.70 & 0.62 & 0.73 & 0.69 & 0.66 & 0.67 & 0.68 & 0.73  \\
\hline
\hline
{Blocks}  &{M$_{11} $}& {M$_{12} $} & {M$_{13} $} & {M$_{14} $} & {M$_{15} $} & {M$_{16} $} & {M$_{17} $} & {M$_{18} $} & {M$_{19} $} & {M$_{20} $} & {M$_{21} $}\\
\hline
Threshold & 0.74 & 0.70 & 0.61 & 0.71 & 0.69 & 0.69 & 0.68 & 0.79 & 0.65 & 0.80 & 0.72\\
\hline
\end{tabular}}
\vspace{-1.5em}
\end{center}

\small
\renewcommand\arraystretch{1.0}
\centering
\setlength{\belowcaptionskip}{0.1cm}
\caption{Offline Evaluation Results on the Twitter dataset. The best results are in bold, and the second-best are underlined.}
\label{table: Offline Evaluation Twitter}
\setlength{\tabcolsep}{2mm}{
\begin{tabular}{c|cccccccccccc}
\hline
Metrics &  Word2vec&  LDA     & WMD     &  BERT   &  BiLSTM  &  PP-GCN  & EventX  &  KPGNN  & QSGNN$^*$ &  HCRC$_{NMI}$   & $\Delta$  \\ \hline
NMI     & .44±.00 & .29±.00 & .65±.00 & .64±.00 & .63±.00 & .68±.02 & \underline{.72±.00}  & .70±.01 & .69±.01 & \textbf{.75±.02} & $\uparrow$ 3\% \\
ARI    & .02±.00  & .01±.00 & .06±.00 & .07±.00 & .17±.00 & .20±01  & .05±.00 & .22±.01 & \underline{.25±.02} & \textbf{.37±.01} & $\uparrow$ 12\% \\ \hline
\end{tabular}}
\vspace{-0.8em}
\small
\setlength{\belowcaptionskip}{0.1cm}
\caption{Offline Evaluation Results on the MAVEN dataset. The best results are in bold, and the second-best are underlined.}
\label{table: Offline Evaluation MAVEN}
\setlength{\tabcolsep}{2mm}{
\begin{tabular}{c|ccccccccccc}
\hline
Metrics &  Word2vec&  LDA     & WMD     &  BERT   &  BiLSTM  &  PP-GCN  & EventX  &  KPGNN   & QSGNN$^*$ &  HCRC$_{NMI}$ & $\Delta$  \\ \hline
NMI     & .42±.00 & .35±.00 & .46±.00 & .45±.00 & .44±.00 & .49±.01 & \underline{.69±.00} & .52±.01 & .55±.03  & \textbf{.70±.03}  &   $\uparrow$ 1\%  \\
ARI     & .02±.00 & .01±.00 & .04±.00 & .02±.00 & .02±.00 & .06±.00 & .00±.00 & \underline{.10±.00} &  .09±.02 &\textbf{.13±.02}   &  $\uparrow$ 3\%  \\ \hline
\end{tabular}}

\end{table*}

\subsection{Incremental Evaluation} 
\label{sec:increexp}
\subsubsection{Traditional Incremental Clustering}
\label{sec:tradintrc}
Since the clustering method in KPGNN and QSGNN is K-Means, the ground-truth label must be required to determine the hyperparameter $k$. For fairness, we compare KPGNN and QSGNN with HCRC$_{NMI}$, and such setting is called \textit{traditional incremental clustering}. 

Table \ref{table: Traditional Incremental evaluation NMIs. }, \ref{table: Traditional incremental evaluation AMIs.} and \ref{table: Traditional incremental evaluation ARIs.} summarize the results. We observe that HCRC$_{NMI} $ achieves the best or second-best performance across most message blocks. HCRC$_{NMI}$ outperforms EventX by 13\% in NMI, 49\% in AMI, and 48\% in ARI, and BERT by 8\% in NMI, 6\% in AMI, and 27\% in ARI on average. This is because EventX only considers community detection and BERT ignores the structural information of social networks. Furthermore, HCRC$_{NMI}$ outperforms KPGNN for most message blocks, because HCRC$_{NMI}$ not only learns the structural information between messages but also effectively learns the semantic information of a single message, but KPGNN only establishes the structural relationship between messages. And, as shown in Table \ref{table: Traditional Incremental evaluation NMIs. }, \ref{table: Traditional incremental evaluation AMIs.} and \ref{table: Traditional incremental evaluation ARIs.}, it is observed that HCRC$_{NMI}$ demonstrates improvements of 2\%, 2\%, and 29\% over QSGNN in NMI, AMI and ARI. However, limited by the experimental environment, we construct multiple social message relation graphs on larger message blocks, such as M$_1$ and M$_7$, rather than a single graph like on other message blocks, leading to lower NMI, but higher ARI. Some message blocks have most messages with similar attributes, resulting in the social message relation graph that is close to a complete graph and causes mediocre clustering performance with HCRC$_{NMI}$, such as with M$_{18}$ and M$_{20}$.

\subsubsection{Semi-Supervised Incremental Clustering}
In practical scenarios, ground-truth labels are difficult to obtain, so the traditional incremental clustering cannot sufficiently fit the real-world setting. To this end, we conduct the semi-supervised incremental clustering, which only provides 10\% available ground-truth labels. The first two rows of Table \ref{table: Semi-supervised and Solid Incremental evaluation NMIs.}, \ref{table: Semi-supervised and Solid Incremental evaluation AMIs.} and \ref{table: Semi-supervised and Solid Incremental evaluation ARIs.} show the experimental results for KPGNN$_{10\%} $, QSGNN$_{10\%} $ and HCRC$_{10\%} $. HCRC$_{10\%} $ outperforms KPGNN$_{10\%} $ by 16\% in NMI, 13\% in AMI and 18\% in ARI on average.  HCRC$_{10\%} $ outperforms QSGNN$_{10\%} $ by 16\% in NMI, 15\% in AMI and 17\% in ARI on average. This is because there are only 10\% available ground-truth labels, and KPGNN$_{10\%} $ and QSGNN$_{10\%} $ cannot get the real number of clusters, which causes K-Means to fail to cluster normally. HCRC$_{10\%} $ can better explore the discriminative information from the social data thereby determining the appropriate thresholds within limited ground-truth labels. However, as demonstrated in traditional incremental clustering shown in Section \ref{sec:tradintrc}, the performance of HCRC in dealing with large message blocks is not satisfactory. This issue is further exacerbated as the amount of available label information decreases, leading to reduced NMI and ARI values for both M$_1$ and M$_7$.

\subsubsection{Solid Incremental Clustering}
To comprehensively evaluate the performance of the proposed HCRC, we further perform solid incremental clustering comparisons, which require that no label information is available. In the last three rows of Table \ref{table: Semi-supervised and Solid Incremental evaluation NMIs.}, \ref{table: Semi-supervised and Solid Incremental evaluation AMIs.} and \ref{table: Semi-supervised and Solid Incremental evaluation ARIs.}, the results of HCRC$_{random}$, KPGNN$_{random} $, QSGNN$_{random} $, and HCRC are provided. The empirical results indicate that on average, HCRC performs better than HCRC$_{random}$ 3\% in NMI, 4\% in AMI and 7\% in ARI, outperforms KPGNN$_{random}$ by 11\% in NMI, 11\% in AMI and 19\% in ARI, and outperforms QSGNN$_{random}$ by 13\% in NMI, 10\% in AMI and 11\% in ARI. The improvements can be attributed to the well-designed DRL-SinglePass, which demonstrates that DRL-SinglePass cannot only break through the limitation of K-Means in the incremental clustering, i.e., requiring the ground-truth label information, but also boost the performance of the clustering model by leveraging the deep reinforcement learning to derive appropriate thresholds. This observation also shows that the proposed HCRC is robust against the negative impact brought by the partial availability of the ground-truth label information. On the contrary, the blocked accessibility of label information excessively degenerates the performance of benchmark methods. To better understand the effectiveness of the proposed DRL-SinglePass, we summarize the derived thresholds for message blocks in Table \ref{table: thresholds}, and the results support that DRL-SinglePass can indeed learn appropriate thresholds for HCRC.

\subsection{Extended Evaluation}

In this subsection, we compare HCRC$_{NMI}$ to other baselines in an offline traditional setting, whereby all datasets are partitioned into training, testing, and validation sets at a ratio of 70\%, 20\%, and 10\%, respectively. The experimental results, as demonstrated in Table~\ref{table: Offline Evaluation Twitter} and~\ref{table: Offline Evaluation MAVEN}, reveal that HCRC$_{NMI}$ outperforms other baselines across all metrics. This is attributed to the fact that baselines such as Word2vec, LDA, WMD, BERT, and BiLSTM disregard the latent structural information in social networks. Furthermore, PP-GCN presumes a stationary graph structure, which is inadequate in capturing dynamic social streams~\cite{KPGNN}. EventX tends to generate more clusters, regardless of whether it captures any additional information or not~\cite{KPGNN}. KPGNN prioritizes structural information over semantic information, as it solely constructs a social network among messages. Although QSGNN primarily focuses on generalizing the model from known data to unknown data, it is similar to KPGNN in that it still places emphasis on structural information. Different from them, HCRC$_{NMI}$ leverages both semantic and structural information in social networks to acquire a more extensive understanding.

Further, we perform the significance test, i.e., t-test, and observe that the P values are consistently lower than 0.05, e.g., 0.012 on the MAVEN dataset. This indicates that the improvement achieved by HCRC is statistically significant, further reinforcing the effectiveness and superiority of HCRC.


\begin{figure*}[t]
	\vskip -0in
	\begin{center}
		\centerline{\includegraphics[width=1.0\textwidth]{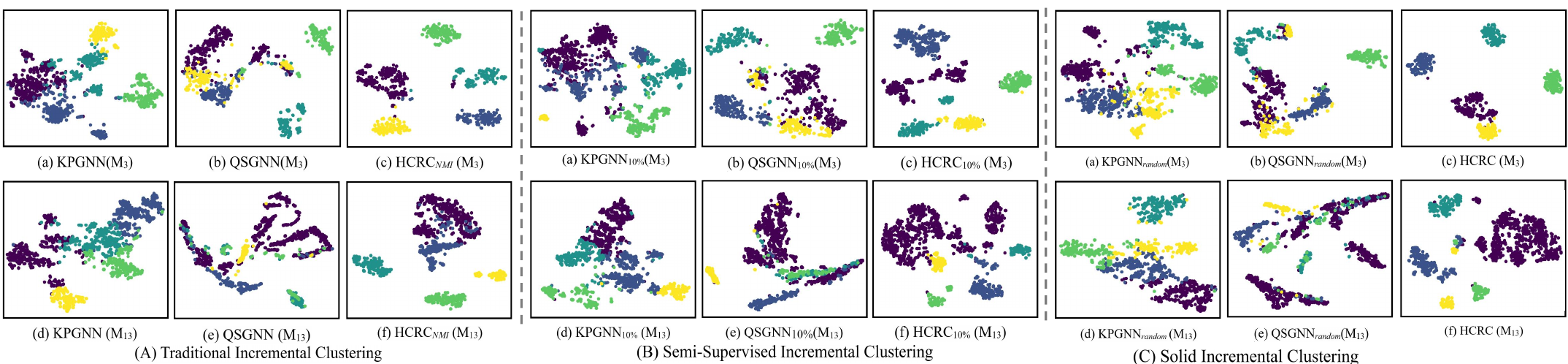}}
		\vskip -0.in
		\caption{T-SNE visualization of the learned message representation on M$_3$ and M$_{13}$.} 
        \label{fig:case}
	\end{center}
\vspace{-1.0em}
\end{figure*}

\begin{table*}
\small
\renewcommand\arraystretch{1.0}
\setlength{\belowcaptionskip}{-0.15cm} 
\caption{Ablation Study. HCRC$_{G-CL}$ and HCRC$_{N-CL}$ represent solely using graph-level contrastive learning and node-level contrastive learning module, respectively, during the model training process. The best results are marked in bold.}
\label{tab: Ablation Study NMIs.}
\begin{center}
\setlength{\tabcolsep}{2mm}{
\begin{tabular}{c|ccccccc}
\hline
{Blocks}  & {M$_{1} $}  & {M$_{2} $} & {M$_{3} $} & {M$_{4} $} & {M$_{5} $} & {M$_{6} $} & {M$_{7} $}  \\
\hline
HCRC & .27±.00 & \textbf{.83±.00} & \textbf{.81±.01} & \textbf{.67±.03} & \textbf{74±.01} & \textbf{.83±.01} & \textbf{.50±.01}\\
HCRC$_{random} $ & \textbf{.27±.01} & .82±.00 & .79±.03 & .63±.05 & .70±.01 & .80±.00 &.46±.05  \\
\hline
HCRC$_{NMI} $ & \textbf{.30±.01} & \textbf{.85±.00} & \textbf{.83±.00} & \textbf{.71±.01} & \textbf{.77±.00} & \textbf{.85±.00}  & \textbf{.52±.00}  \\
HCRC$_{G-CL}$  & .27±.00 & .77±.01 & .73±.01 &	.65±.00 & .65±.01 &	.80±.01 & .46±.01\\
HCRC$_{N-CL}$ &.25±.00 & .77±.01&	.76±.02 & .70±.00&	.70±.00 & .76±.02& .48±.00 \\
\hline
\end{tabular}}

\setlength{\tabcolsep}{2mm}{
\begin{tabular}{c|ccccccc}
\hline
{Blocks}  & {M$_{8} $}  & {M$_{9} $} & {M$_{10} $} & {M$_{11} $} & {M$_{12} $} & {M$_{13} $} & {M$_{14} $}  \\
\hline
HCRC & \textbf{.78±.01} & \textbf{.76±.00} & \textbf{.80±.01} & \textbf{.70±.02} & \textbf{.69±.00} & \textbf{.69±.04} & \textbf{.68±.01} \\
HCRC$_{random} $ & .75±.01 & .70±.01 & .78±.01 & .67±.01 & .65±.01 & .67±.01 & \textbf{.68±.01} \\
\hline
HCRC$_{NMI} $  & \textbf{.81±.00} & \textbf{.79±.00} & \textbf{.84±.00} & \textbf{.80±.00} & \textbf{.70±.01} & \textbf{.79±.00} & \textbf{.71±.01}\\
HCRC$_{G-CL}$ &.77±.02 & .70±.01 & .78±.01 & .64±.00 & .62±.00 & .68±.03 &	.65±.01 \\
HCRC$_{N-CL}$  &.73±.01 & .74±.00 & .74±.00 & .73±.00 & \textbf{.70±.01} & .74±.01& .67±.01\\
\hline
\end{tabular}}

\setlength{\tabcolsep}{2mm}{
\begin{tabular}{c|ccccccc}
\hline
{Blocks}  & {M$_{15} $}  & {M$_{16} $} & {M$_{17} $} & {M$_{18} $} & {M$_{19} $} & {M$_{20} $} & {M$_{21} $}  \\
\hline
HCRC & \textbf{.68±.02} & \textbf{.86±.02} & \textbf{.71±.02} & \textbf{.61±.01} & \textbf{.74±.01} & \textbf{.69±.02} & \textbf{.56±.01}  \\
HCRC$_{random} $ & .58±.01 & .80±.02 &.67±.01 & .60±.01 & .69±.02 & .69±.01 & .55±.03 \\
\hline
HCRC$_{NMI} $ & \textbf{.70±.01} & \textbf{.87±.00} & \textbf{.75±.00} & \textbf{.63±.01} & \textbf{.76±.01} & \textbf{.72±.00} & \textbf{.62±.00} \\
HCRC$_{G-CL}$  & .54±.01 & .82±.03 & .63±.02 & .56±.01 & .65±.00 &	.71±.01 & .57±.01 \\
HCRC$_{N-CL}$ &.67±.01 & .78±.01 & .68±.01& .63±.00 &	.71±.00& .70±.00& .58±.01\\
\hline
\end{tabular}}
\end{center}

\end{table*}

\begin{figure}[t]
	\vskip -0in
	\begin{center}
\centerline{\includegraphics[width=0.5\textwidth]{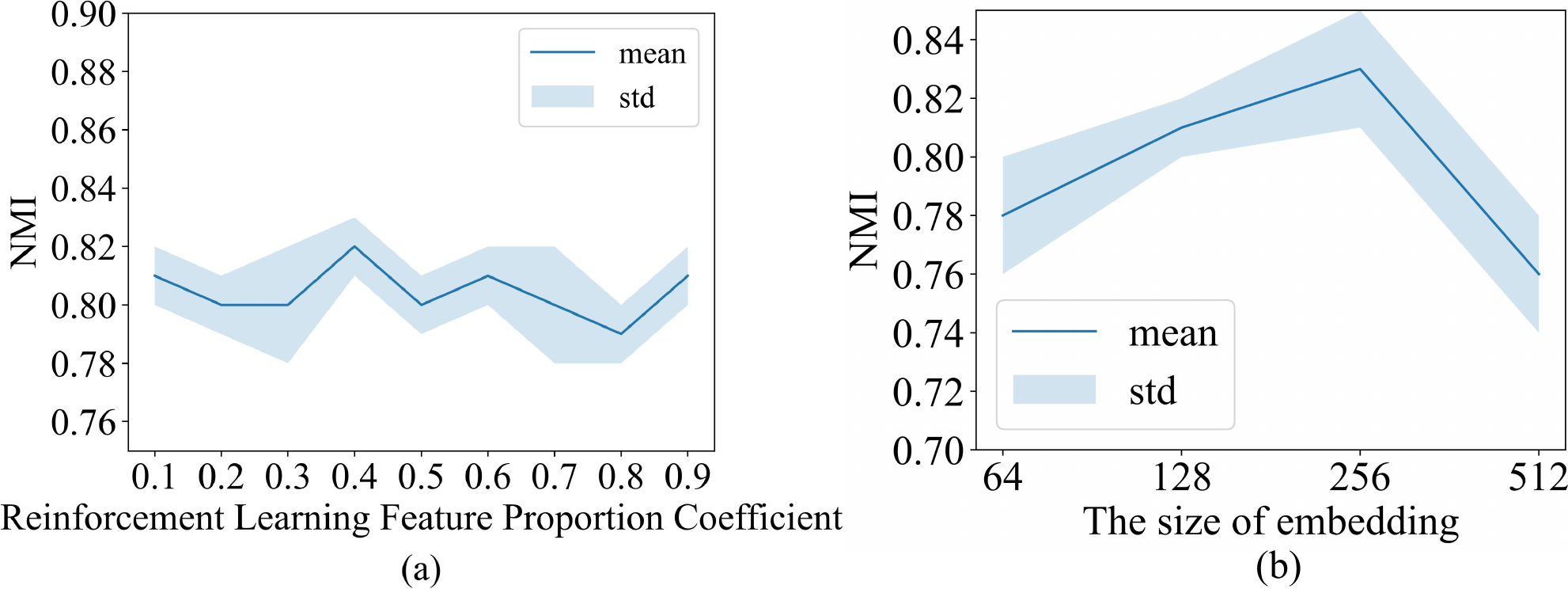}}
		\caption{Analysis of Hyperparameter.} 
        \label{fig:hyper}
	\end{center}
\vspace{-2.0em}
\end{figure}
\subsection{Visualization Results}
In this subsection, we use t-Distributed Stochastic Neighbor Embedding (T-SNE) \cite{van2008visualizing} to reduce the dimensionality of the message representation in M$_3$ and M$_{13}$ to two dimensions. We further present visualizations of the clustering outcomes obtained from KPGNN, QSGNN and HCRC across three distinct experimental settings, aiming to provide additional evidence of the superiority of HCRC. Our attention is predominantly directed towards the five most prevalent events in terms of tweet volume, taking into account the long-tail challenge prevalent in social data. Tweets pertaining to the same event are represented using consistent color markers. The results depicted in Fig. \ref{fig:case} provide compelling evidence that HCRC consistently outperforms both KPGNN and QSGNN in terms of producing a more compact clustering outcome with clearly defined boundaries. This superiority of HCRC holds true across all experimental settings, indicating its robustness and effectiveness across all experimental settings. HCRC clearly achieves superior performance and demonstrates greater adaptability to incremental event detection.

\subsection{Analysis of Hyperparameter}
In this section, we analyze two crucial hyperparameters, the reinforcement learning feature proportion coefficient and the size of embedding. To explore their sensitivity, we conduct a comprehensive evaluation of the model's performance on message block M$_3$.

The reinforcement learning feature proportion coefficient is used in DRL-SinglePass to learn data features from how much proportion of tweets in order to obtain an appropriate threshold. As illustrated in Fig. \ref{fig:hyper} (a), the model's performance demonstrates minimal fluctuations when adjusting the proportion coefficient, signifying its insensitivity to this particular hyperparameter. In light of practical considerations and the need to optimize training efficiency, we have chosen to compromise a marginal fraction of the model's performance by setting the reinforcement learning feature proportion coefficient to 0.1. Based on Fig. \ref{fig:hyper} (b), we observed that within the range of (64, 128, 256, 512), the model demonstrates the best performance as the embedding size increases. When the embedding size is set to 256, although there is an increase in computational complexity, HCRC can capture the tweets' more semantic information and exhibit better discriminative power. Therefore, we decide to set the embedding size to 256. 

\subsection{Ablation Study}
In this subsection, we conduct the ablation study on HCRC using NMI, analyzing the effectiveness of its constituent components, and the comparisons are shown in Table \ref{tab: Ablation Study NMIs.}. From the first two rows of Table \ref{tab: Ablation Study NMIs.}, it can be seen that in the absence of any available label information, DRL-SinglePass demonstrates its effectiveness in social event detection. From the last three rows of Table \ref{tab: Ablation Study NMIs.}, compared with HCRC$_{NMI}$, the variants eliminating either graph-level contrastive learning (G-CL) or node-level contrastive learning (N-CL) generally underperform the complete model, which demonstrates the effectiveness of the proposed graph-level and node-level contrastive learning. From the results in Table \ref{table: Traditional Incremental evaluation NMIs. } and Table \ref{tab: Ablation Study NMIs.}, we conclude that although KPGNN generally beats both the HCRC variants without G-CL or N-CL, the complete HCRC can outperform KPGNN. The remarkable performance boost verifies the superiority of hybrid graph contrastive learning.

\section{Conclusion and Future Work}
We clarify the issues existing in benchmark methods, i.e., the adopted GCL cannot sufficiently capture the semantic information of social messages. Current embedding clustering approaches exceptionally adopt the data-related information resulting in the breach of the solidly unsupervised warranty. To this end, we propose HCRC to learn the comprehensive semantic and structural information from social messages by using hybrid graph contrastive learning, and the proposed reinforced incremental clustering empowers HCRC to perform solid incremental clustering. Empirically, HCRC outperforms baselines in various experimental settings.

Due to the incremental nature of HCRC, we do not impose any practical restrictions on the dataset size, such that the Twitter dataset used in our experiments can be continuously maintained, thereby enabling the extension of HCRC to larger datasets. Due to the limitation of the available datasets, we can only perform the evaluation of our approach on the adopted datasets that meet the required criteria. Therefore, in future work, our principal emphasis will center on exploring the application of HCRC to large-scale datasets.

\section*{Acknowledgments}
The authors would like to thank the anonymous reviewers for their valuable comments. This work is supported by the Fundamental Research Program, Grant No. JCKY2022130C020, and the Strategic Priority Research Program of the Chinese Academy of Sciences, Grant No. XDA19020500.
\bibliographystyle{elsarticle-num} 
\bibliography{references}


\end{document}